\documentclass[superscriptaddress,amsmath,amssymb]{revtex4-2}

\usepackage[english]{babel}
\usepackage[T1]{fontenc} 
\usepackage{lmodern} 
\usepackage{xcolor}
\usepackage{miller}
\usepackage{caption}
\usepackage{subcaption}
\usepackage{ragged2e}

\DeclareCaptionFormat{justified}{\justifying#1#2#3}
\usepackage{graphicx}
\usepackage{textcomp}

\makeatletter
\def\maketitle{
\@author@finish
\title@column\titleblock@produce
\suppressfloats[t]}
\makeatother

\def \FUW{Faculty of Physics, University of Warsaw, Pasteura St. 5, 02-093 Warsaw, Poland}
\def \UNIPRESS{Institute of High Pressure Physics Polish Academy of Sciences, 29/37 Sokolowska St., 01-142 Warsaw, Poland}
\def \CNT{CENTERA, CEZAMAT, Warsaw University of Technology, 19 Poleczki Str., 02-822 Warsaw, Poland}
\def \PL{Institute of Physics, Łódź University of Technology, 217/221 Wólczańska St., 90-451 Łódź, Poland}
\def \IFPAN{Institute of Physics, Polish Academy of Sciences, 32/46 Lotnikow Av., 02-668 Warsaw, Poland}
\def \mesco{Center of Development and Implementation, Telesystem-Mesko Sp. z o.o, ul. Warszawska 51, 05-082, Lubiczów, Poland}
\def \cezamat{CEZAMAT, Warsaw University of Technology, Poleczki 19, 02-822, Warsaw, Poland}
\begin{document}
\title{Optical bound states in the continuum in subwavelength gratings made of an epitaxial van der Waals material}
\author{Emilia Pruszyńska-Karbownik}\affiliation{\FUW}
\email{emilia.karbownik@fuw.edu.pl}
\author{Tomasz Fąs}\affiliation{\FUW}
\author{Katarzyna Brańko}\affiliation{\FUW}
\author{Dmitriy Yavorskiy}\affiliation{\UNIPRESS}
\affiliation{\IFPAN}
\affiliation{\CNT}
\author{Bartłomiej Stonio}\affiliation{\cezamat}
\author{Rafał Bożek}\affiliation{\FUW}
\author{Piotr Karbownik}\affiliation{\mesco}
\author{Jerzy Wróbel}\affiliation{\IFPAN}
\author{Tomasz Czyszanowski}\affiliation{\PL}
\author{Tomasz Stefaniuk}\affiliation{\FUW}
\author{Wojciech Pacuski}\affiliation{\FUW}
\author{Jan Suffczyński}\affiliation{\FUW}

\date{\today}


\maketitle

\keywords{bound state in the continuum, van der Waals material, transition metal dichalcogenides, molybdenum diselenide, subwavelength grating, molecular beam epitaxy, third harmonic generation}
\begin{abstract}
High refractive index (4.4 at 1100 nm), negligibly small absorption in near-infrared spectral range, and ease of processing make MoSe$_2$ a perfect material for applications in near-infrared photonics. So far, implementation of MoSe$_2$-based photonic structures has been hindered by the lack of large surface MoSe$_2$ substrates. The use of molecular beam epitaxy allows the production of homogeneous layers of MoSe$_2$ with a few-inch surface and a thickness controlled at the sub-nm level. In the present work, we design by theoretical calculations and fabricate by a simple lithography process an ultrathin subwavelength grating out of 42-nm thick, epitaxially-grown MoSe$_2$ layer. Our polarization-resolved reflectivity measurements confirm that the gratings host a peculiar type of a confined optical mode that is a bound state in the continuum. Moreover, the fabricated structures enhance the efficiency of the third harmonic generation by over three orders of magnitude as compared to the unstructured MoSe$_2$ layer. The presented results are promising for the realization of flat, ultra-compact devices for lasing, wavefront control, and higher-order topological states of the light.

\end{abstract} 

\section*{Introduction}
An optical bound state in the continuum (BIC) represents a peculiar type of a non-radiating electromagnetic resonant state that, despite being spatially confined in an open photonic system, coexists with a continuous spectrum of unbounded states (above the light line) \cite{Marinica2008PRL,Plotnik2011PRL}. A BIC ensures the confinement of the light to subwavelength dimensions with the corresponding sharp optical resonance characterized by an infinite quality factor. Unique optical properties of the BICs are advantageous for diverse device applications such as low-threshold nano lasers,\cite{Robbiano2018AN} efficient single photon sources,\cite{Wang2019NPh} and fundamental studies including Bose–Einstein condensation\cite{Hakala2018NP} or superfluidity\cite{He2020NP}.

Realization of optical BICs requires a spatial modulation of the refractive index in a subwavelength scale, leading to a plane wave coupling in the direction perpendicular to the structure and a waveguiding effect in its plane \cite{Glowadzka2021NPh}. Typically, such modulation is achieved through a periodic pattering of a thin layer of a dielectric material, whose refractive index is significantly larger than that of the surroundings, particularly of the substrate it is deposited on. With the nano-structuration, however, the average refractive index of the patterned layer decreases, which may hinder the formation of BICs. Therefore, searching for materials with a high refractive index that could form a periodic structure is of great importance for practical applications of BIC-based cavities. 
\begin{figure}
    \centering
    \includegraphics[width=0.5\textwidth]{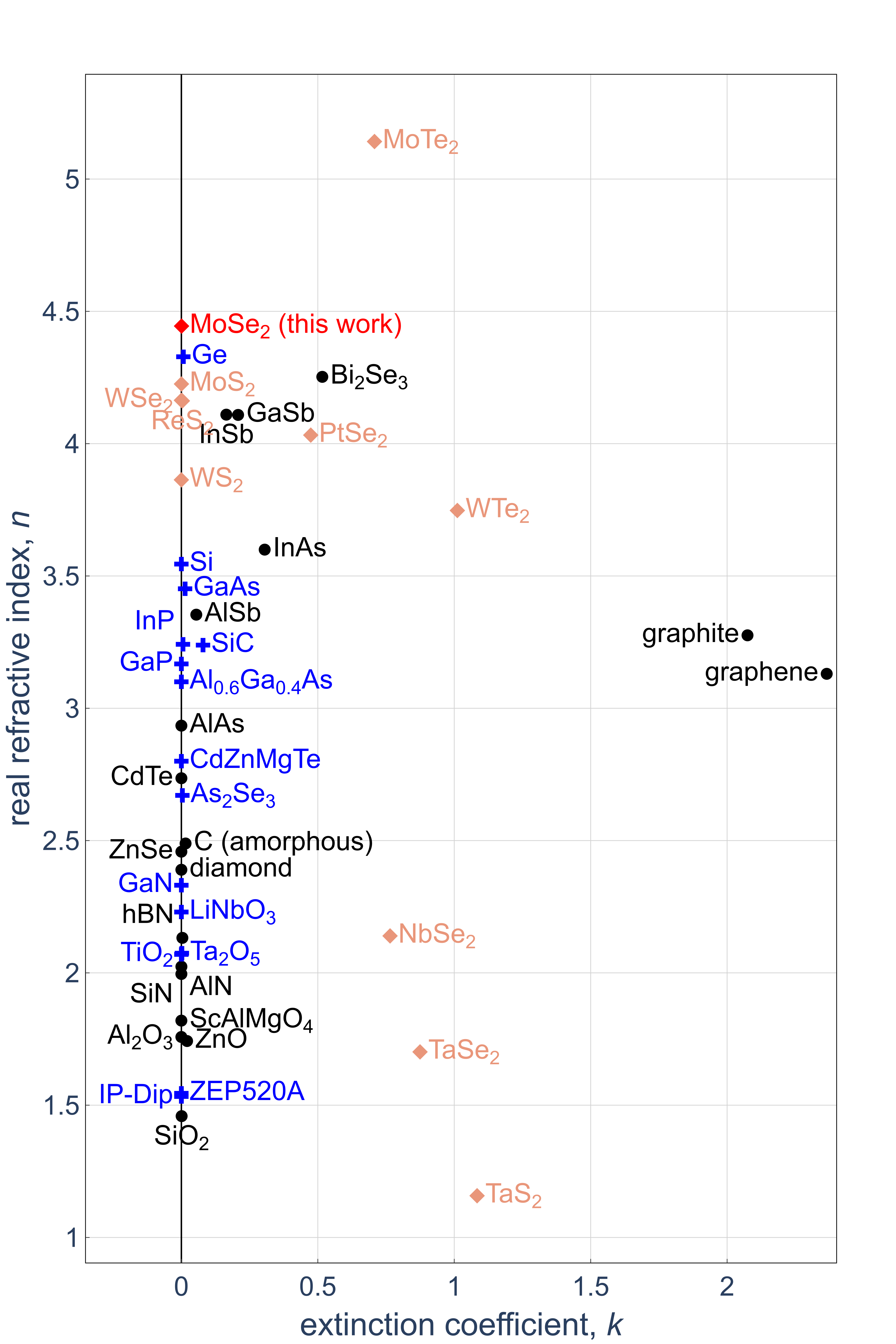}
    \caption{Real $n$ and imaginary $k$ part of refractive indices for light at the wavelength of $\lambda=1100\,\mathrm{nm}$ for III-V semiconductors: GaAs, AlGaAs, InAs, InP, GaP, InSb, AlSb \cite{Adachi1989JAP}, AlAs \cite{Rakic1996JAP}, GaN \cite{Barker1973PRB}, AlN \cite{Beliaev2021JVST}, GaSb \cite{Ferrini1998JAP}, group IV semiconductors: Si \cite{Schinke2015AIPA}, Ge \cite{Amotchkina2020AO}, SiC \cite{Larruquert2011JOSAA}, TiC \cite{Pfluger1984PRB}, diamond \cite{Philip1964PR}, amorfous carbon \cite{Larruquert2013OE}, graphite \cite{Djurisix1999JAP}, graphene \cite{Tikuisis2023PRM}, oxides:  SiO$_2$ \cite{Rodriguez2016OME}, TiO$_2$ \cite{Sarkar2019ACSAMI}, Ta$_2$O$_5$ \cite{Bright2013JAP}, Al$_2$O$_3$ [Fig. S1 in the Supporting Information], ZnO \cite{Aguilar2019OME}, LiNbO3 \cite{Zelman1997JOSAB}, ScAlMgO$_4$ \cite{Stefaniuk2023PRB}, nitrides: SiN \cite{Beliaev2022TSF}, TiN, VN \cite{Pfluger1984PRB}, hexagonal BN \cite{Grudinin2023MH}, selenides: As$_2$Se$_3$ \cite{Joseph2020ACSAMI}, Bi$_2$Se$_3$ \cite{Fang2020ASS}, ZnSe \cite{Querry1987CR}, tellurides: CdTe \cite{Treharne2011JPCS}, resins: IP-Dip \cite{Gissibl2017OME}, ZEP520A \cite{ZEONreport}, and transition metal dichalcogenides (indicated by diamonds in redish colors): MoS$_2$, MoTe$_2$, WS$_2$, WSe$_2$, WTe$_2$, NbSe$_2$, ReS$_2$, TaS$_2$, TaSe$_2$ \cite{Munkhbat2022ACSPh}, PtSe$_2$ \cite{Ermolaev2021Nm}, MoSe$_2$ [Fig.~\ref{fig:techn}c herein]. See also \cite{Polyanskiy2024SD}. Materials already used as subwavelength grating material are indicated by blue crosses.}
    \label{fig:nref1100}
\end{figure}

Subwavelength high-refractive-index-contrast gratings\cite{Mateus2007IEEEPTL} -- dielectric diffraction gratings with a period smaller than the wavelength of the incident light -- are renowned for their ability to host BICs. Various material systems have been considered in search of an optimal grating material, in particular, III-V semiconductors \cite{Zhou2008OE,Chase2010OE,Ansbaek2013IEEEPTL,Zhang2020SSC,Marciniak2021ACSP}, group IV semiconductors \cite{Lai2015procSPIE,Hogan2016OL}, oxides \cite{Hashemi2015JVST,Benimetskiy2020JPCS}, II-VI semiconductors \cite{Marciniak2021ACSP}, selenides \cite{Joseph2021AOM}, and resins \cite{Wang2021ACSN,EPK2023NPh}. The optimal candidate needs to meet, among others, three requirements: zero or very low absorption, as high as possible real part of the refractive index, and ease of fabrication and nanostructuration. Figure~\ref{fig:nref1100} shows the real and imaginary parts of the refractive index at the wavelength of 1100 nm for selected semiconductors or dielectrics originating from different material classes. In the near-infrared range of the spectrum, the above-listed requirements for the refractive index are fulfilled perfectly by molybdenum diselenide (MoSe$_2$) -- a prominent member of the transition metal dichalcogenide (TMD) family \cite{Duong2017ACSN}. In addition, the technology of nano- or microstructuration of the MoSe$_2$ is simple and already well developed \cite{Ting:AMT2019}.


An exceptionally high refractive index (of up to 4.5 -- 5) is a remarkable property of MoSe$_2$ that still has been mostly out of the focus of the researchers \cite{Duong2017ACSN,Singh:ACSPhotonics2020,Lin2020LSA}. So far, van der Waals materials have been employed in subwavelength structures for light guiding in a photonic crystal made of mono- or a few-layer WS$_2$ \cite{Zhang2019NN} and for atomically thin lenses made of a WSe$_2$ monolayer \cite{Lin2020LSA, Guarneri2024NL}. The main obstacle to the implementation of the TMD materials in the light-confining structures has been the lack of large, homogeneous TMD layers with tens of nanometers thickness. 

To date, no experimental reports of subwavelength gratings made of any TMD semiconductor and no BIC-type optical states have been demonstrated in such gratings so far. There have been only theoretical predictions of BICs in photonic structures made of WS$_2$ layers,\cite{Zong2021OL,Qin2023PRB} and very recently 2D photonic metasurfaces hosted by nanostructured TMD flakes obtained by exfoliation\cite{Isoniemi:ACSNano2024,Maggiolini:NatureMat2023} have been reported. In parallel, subwavelength dielectric resonators hosting BICs in conventional semiconductors such, as (Al,Ga)As, GaP or SiN,\cite{Ning2020OE,Shcherbakov2021NC,Qiu2023OC} have been employed to boost the light-matter coupling and generation of higher-order harmonics in unstructured van der Waals monolayers.\cite{Wang:Nanophotonics2024, Lee:NanoLett2025} 

In this work, firstly, we theoretically predict the optical BIC states in subwavelength grating structures made of a few-tens-of-nm thick layer of MoSe$_2$. Next, we present a sample design and we employ molecular beam epitaxy (MBE) to fabricate a few-inch large, homogeneous layers of MoSe$_2$ of a controlled thickness. We fabricate one-dimensional subwavelength gratings out of the prepared MoSe$_2$ layers using a basic three-step process: photoresist deposition, e-beam lithography, and dry etching, without metal layer deposition and lift-off steps. Our angle-resolved reflectivity measurements on the obtained MoSe$_2$-based subwavelength gratings reveal the optical BIC state confined in the grating at the NIR spectral region. The observation of a polarization vortex in the spectral and in-plane photon momentum vicinity of the $k = 0$ further confirms the presence of the BIC. Finally, we demonstrate that the excitation of an optical mode in spectral proximity to the BIC significantly enhances third harmonic (TH) generation directly in the grating volume, achieving a boost for over three orders of magnitude compared to a bare MoSe$_2$ layer.

\section*{Results and discussion}

\textbf{Design of the MoSe$_2$ gratings}

\begin{figure}[htp]
    \centering
    \includegraphics[width=0.6\linewidth]{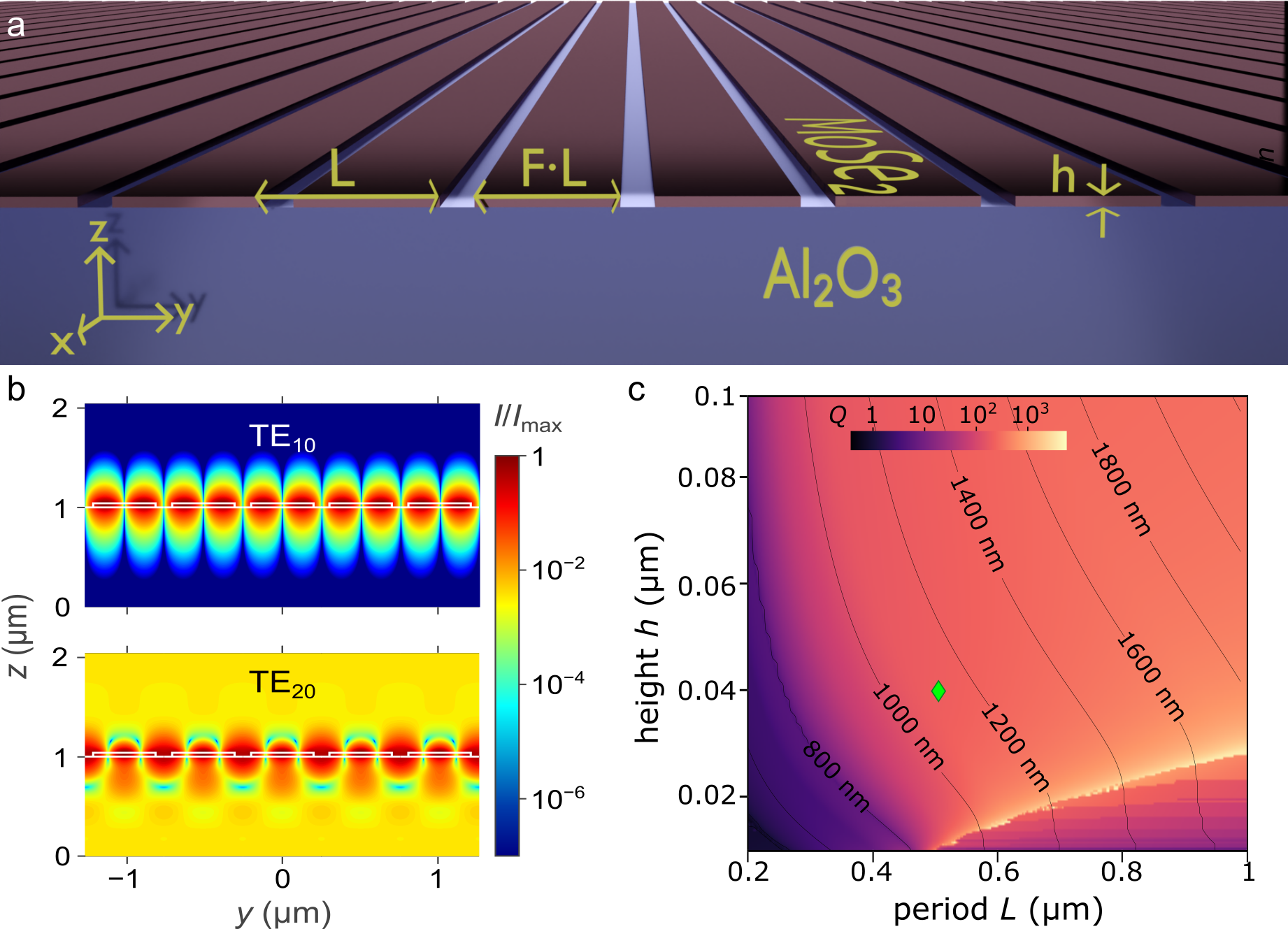}
    \caption{(a) Schematic (in scale) cross-section of a subwavelength grating made of MoSe$_2$ deposited on an Al$_2$O$_3$ substrate. Geometry of the grating is defined by its period $L$, height $h$, and a stripe width $F\cdot L$, where $F$ is a fill factor; 
    (b) Cross-sections of the light intensity distributions in five stripes of the MoSe$_2$ subwavelength grating with the height $h=42\,\mathrm{nm}$, period length $L=500\,\mathrm{nm}$, and fill factor $F=0.8$ for modes TE$_{10}$ and TE$_{20}$. The boundaries of MoSe$_2$ and sapphire layers cross section are indicated with white lines; (c) Numerically calculated map of Q-factor of the optical TE$_{10}$ mode of the MoSe$_2$ subwavelength grating in the space of the grating period and height for the selected fill-factor $F=0.8$ and with the experimentally determined, net absorption of the MoSe$_2$ layer. Contour lines mark the selected wavelengths of the mode. The green diamond indicates parameters of the grating structure selected for the experimental realization.}
    \label{fig:design_all}
\end{figure}

A subwavelength grating geometry is defined by the grating height $h$, period $L$, and fill factor $F$ being the ratio of the width of individual stripe to the period (see Fig.~\ref{fig:design_all}a), as well as by the refractive indices of the grating material and of the surroundings, particularly the substrate.  

We design the MoSe$_2$ subwavelength gratings using numerical calculations. We consider a single period of the grating with periodic boundary conditions in the $y$ direction, infinite length in the $x$ direction, sandwiched between infinitely high layers of air and sapphire in the $z$ direction (see Figure~\ref{fig:design_all}a). Our analysis focuses on the wavelength and Q-factor of the lowest-order optical mode (TE$_{10}$) of the grating, as this anti-symmetric mode can potentially host a BIC. 

The Q-factor is inversely proportional to the cavity's power loss, being the sum of radiation losses and internal absorption. Since in our calculations we consider gratings with an infinite number of periods, there are no radiation-type losses in horizontal directions (the grating plane). Therefore, $\frac{1}{Q}=\frac{1}{Q_{Rz}}+\frac{1}{Q_{abs}}$, where $Q_{Rz}$ accounts for radiation losses in $z$ direction, while $Q_{abs}$ accounts for absorption losses. The optical mode is regarded as the BIC if its radiation losses are determined to be zero, which corresponds to infinitely large $Q_{Rz}$.  

\begin{figure}[htp]
    \centering
    \includegraphics[width=0.53\linewidth]{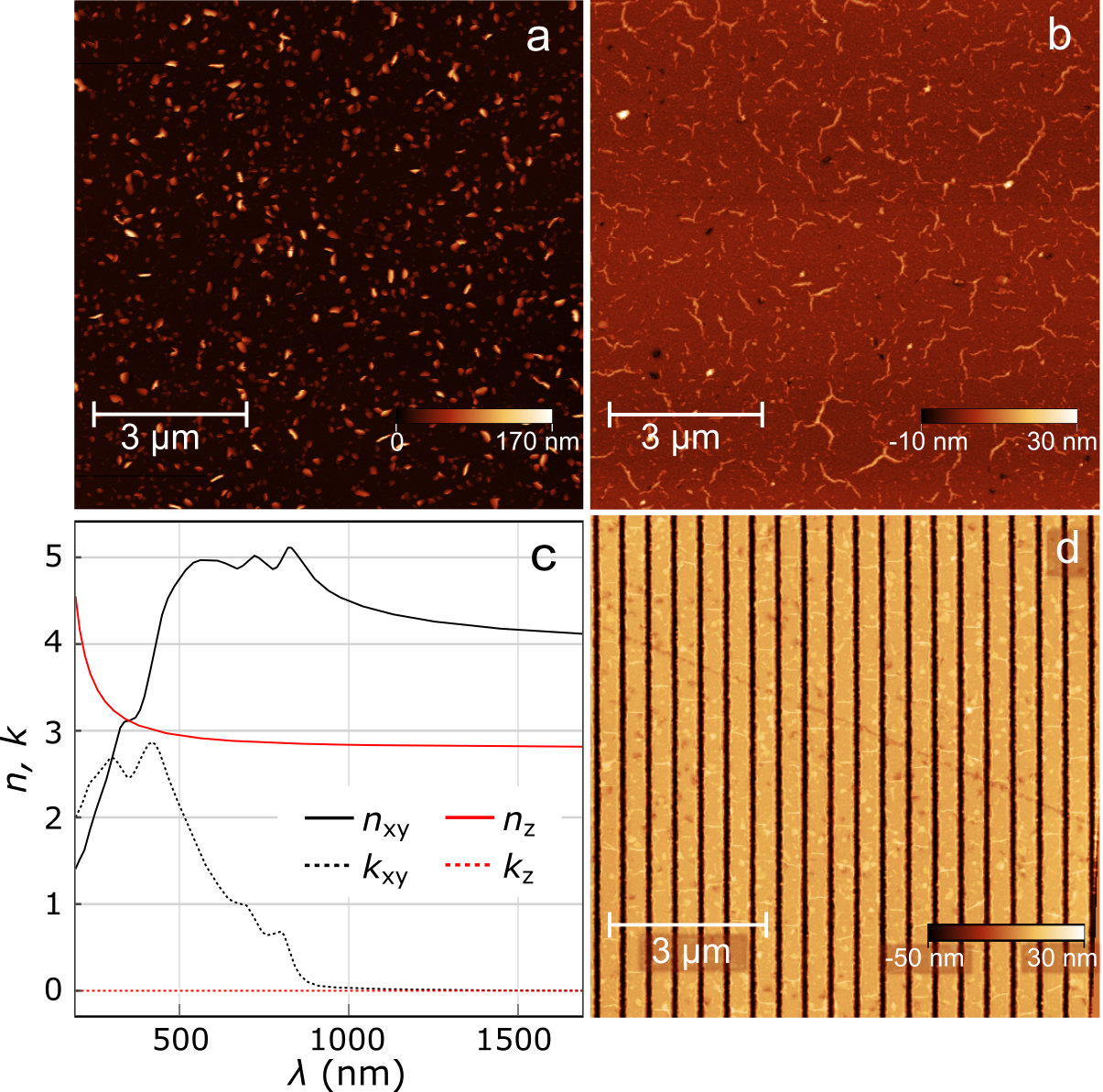}
    \caption{ (a) AFM image of as-grown MoSe$_{2}$ layer surface, (b) AFM image of the MoSe$_{2}$ surface after polishing, (c) Real and imaginary part of in-plane $n_{xy}$ and out-of-plane refractive index $n_{z}$ of the 42-nm MoSe$_2$ layer determined by our ellipsometry measurements, (d) AFM image of the excerpt of the MoSe$_{2}$-based subwavelength grating with the height $h = 42$~nm, period $L=500$~nm and fill factor $F=0.79$.} 
    \label{fig:techn}
\end{figure}

Values of the calculated Q-factor and wavelengths for the TE$_{10}$ mode as a function of the height $h$ and period $L$ for fill factor $F=0.8$ of the MoSe$_2$-based subwavelength grating are presented in Fig.~\ref{fig:design_all}c. Realistic, determined by us in the ellipsometry, complex refractive indices of MoSe$_2$ and sapphire (Fig. \ref{fig:techn}c and Fig S1, respectively) are utilized in the calculations. In view of our calculations, the wavelength $\lambda$ of the optical mode of the grating depends mainly on the period $L$ and only slightly on the height $h$. If the height $h$ of the grating exceeds the cut-off value $h_{\min}$ and when neglecting imaginary part of the MoSe$_2$ refractive index, the quality factor $Q_{Rz}$ of the mode exceeds $10^{12}$, what is a numerical equivalent of the infinity (see also Figure~S2 presenting respective maps for $F=0.5$). In the realistic case the quality factor of the grating's optical mode is never infinite due to a net value of the imaginary part of the refractive index of the grating material, in our case MoSe$_2$ (see Figured~\ref{fig:design_all}c, S2b, and S4). In the following, however, we show that non-negligible absorption of the grating material does not preclude the formation of the BIC. 

Through our calculations, we find that if the grating height $h$ exceeds the cut-off value, the mode becomes non-radiating and can be identified as a BIC. Otherwise, the mode leaks into the substrate hindering the BIC formation. Taking this into account we choose following parameters of the grating: $h=40\,\mathrm{nm}$, $L=500\,\mathrm{nm}$, and fill factor $F=0.8$ (as marked in the Fig.~\ref{fig:design_all}c by a green diamond). The chosen $h$ is high enough to avoid the leakage of the mode and still low enough to make the layer epitaxy and the processing feasible. The $L$ and $F$ are adjusted to ensure that the expected wavelength of the BIC is around 1100~nm and, therefore, remains in the range of possibly low absorption of MoSe$_2$.
The numerically calculated cross-sections of a distribution of the light intensity for the antisymmetric TE$_{10}$ mode for these parameters confirm strong mode confinement in the grating plane and its non-radiating character (see Figure~\ref{fig:design_all}b). For comparison, the symmetric TE$_{20}$ mode is found to strongly leak to the substrate, which precludes the BIC. The calculated distributions of the light intensity for various heights of the grating $h$ are shown additionally in Figure~S2a in the Supporting Information.

\textbf{Sample growth and processing}

Large (of the order of a few cm$^2$) homogeneous MoSe$_2$ layers are grown by MBE. The growth is performed at a very small growth rate of $\sim$ 1 monolayer per hour, proved previously to ensure a high optical-quality of the MoSe$_2$ monolayers\cite{Pacuski2020NL} (see Methods for details). The reflectivity and transmission of the epitaxial layer are shown in Fig. S3. Fig.~\ref{fig:techn}a shows an Atomic Force Microscopy (AFM) image of the surface of the as-grown MoSe$_2$ layer. Due to the tendency to island formation and vertical growth of multilayer TMDs,\cite{Kong2013NL,Li2016SR} and despite the repetitive in-situ annealing steps introduced during the growth, some density of MoSe$_2$ nanopillars is present on the layer surface. To remove the nanopillars, mechanical polishing using a silk tissue is performed. The surface of the same sample after the mechanical polishing is shown in Fig.~\ref{fig:techn}b. The polishing diminishes the surface roughness from $S_q=14.6\,\mathrm{nm}$ for the as-grown layer to $S_q=2.0\,\mathrm{nm}$ for the polished layer.

Ellipsometry measurements are conducted on polished MoSe$_2$ layers with a thickness of $h = 42\,\mathrm{nm}$, to determine their real and imaginary parts of in-plane ($n_{xy}$) and out-of-plane ($n_z$) refractive index.  
We find that the real part of the $n_{xy}$ attains 4.5, while the real part of the $n_z$ is around 3 in the spectral vicinity of 1100~nm (in Fig.~\ref{fig:techn}c). The imaginary parts of the $n_{xy}$ and $n_z$ vanish in the NIR spectral region, in agreement with previous reports \cite{Munkhbat2022ACSPh}.


Subwavelength grating structures are fabricated from homogeneous MoSe$_2$ layers employing e-beam lithography and dry etching. An AFM image showing a part of an example MoSe$_{2}$-based subwavelength grating with period $L=500$~nm and fill factor $F=0.79$ is shown in Fig.~\ref{fig:techn}d. A photograph of the sub-mm area of the sample with an array of 100 \textmu m $\times$ 100 \textmu m gratings is shown in Figure~S6 in the Supporting Information.

\begin{figure}[htp]
    \includegraphics[width=0.7\textwidth]{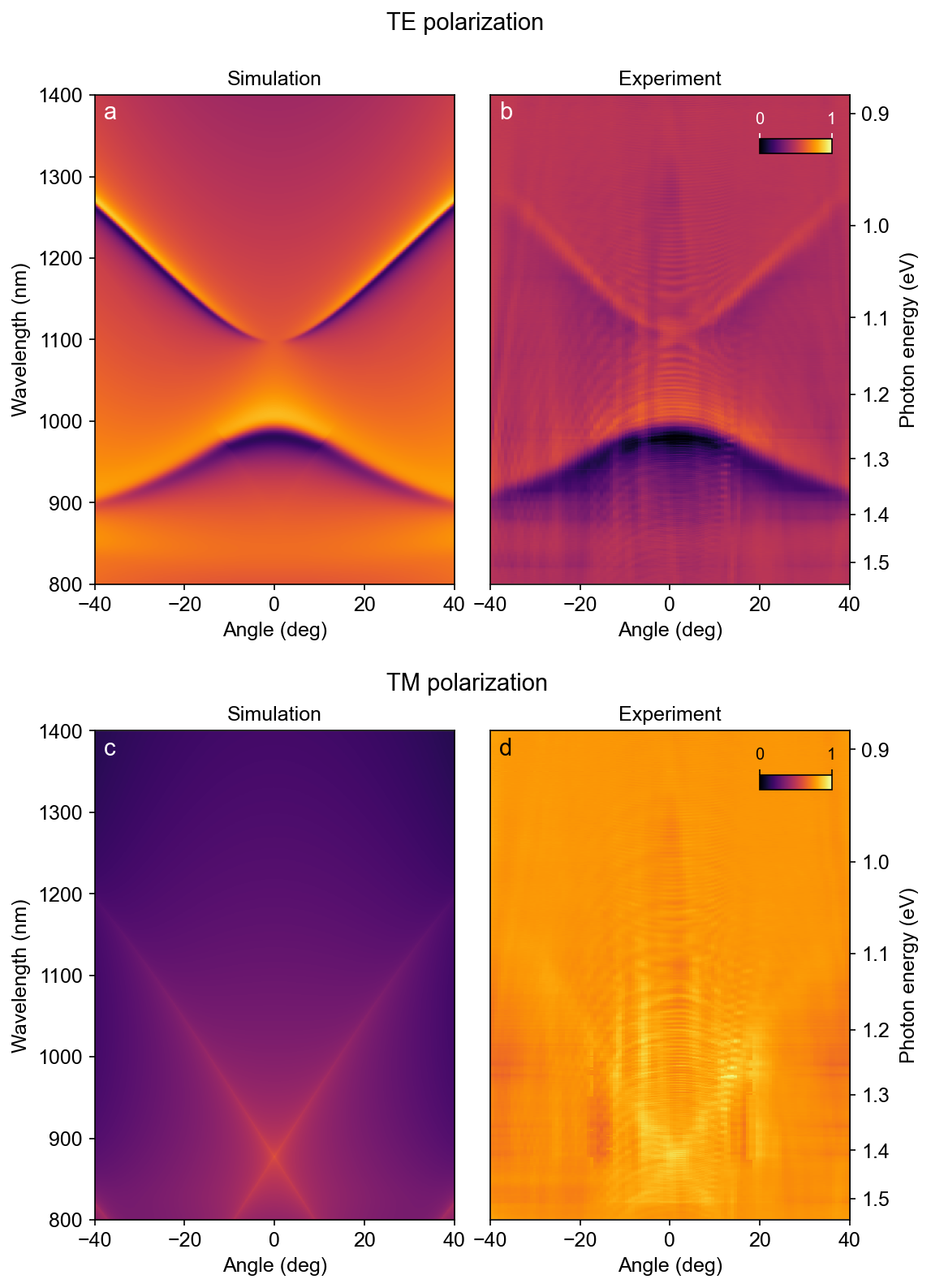}
    \caption{\label{fig:reflectivity_results} Angle-resolved maps of reflectivity from the MoSe$_2$-based subwavelength grating for TE polarization: a) calculated and b) obtained in the experiment, and for TM polarization: c) calculated and d) obtained in the experiment. BIC state is evidenced in TE polarization at around 1100 nm. The geometry of the grating is: height $h=42\,\mathrm{nm}$, period $L=500\,\mathrm{nm}$, and fill-factor $F=0.79$. Colorscale ranges from purple for zero reflectivity to yellow for maximal reflectivity, and the same colorscale is used for the four maps.}
\end{figure}

\textbf{Photon in-plane momentum resolved reflectivity}

The reflectivity spectra of the MoSe$_2$ subwavelength grating calculated as a function of the photon in-plane momentum (represented by the angle) are shown in Fig.~\ref{fig:reflectivity_results}a and Fig.~\ref{fig:reflectivity_results}c for TE and TM polarizations, respectively. Corresponding experimental spectra are shown respectively in Fig.~\ref{fig:reflectivity_results}b and Fig.~\ref{fig:reflectivity_results}d. 
The calculations use the refractive index of MoSe$_2$ obtained in our ellipsometry measurements (see Fig.~\ref{fig:techn}c) and geometry parameters determined by AFM without any additional parameter tuning. A perfect agreement between the experimental data and model description is evidenced.

There are two modes present in the reflectivity map acquired in TE polarization. Both show a Fano-type shape, as indicated by an abrupt change from a high to a low value when crossing the mode central wavelength. The mode at the wavelength of around 980~nm at $k = 0$ exhibits a standard (positive) dispersion, with the energy increase with the increasing in-plane photon momentum wavevector $k$ (or angle). Numerical calculations identify this mode as the symmetric TE$_{20}$ mode. The mode with a minimum at 1100 nm at $k = 0$ shows an anomalous dispersion, which indicates that it is antisymmetric with respect to the cross-section of the grating. The calculations identify it as the TE$_{10}$ mode. Upon approaching $k = 0$ the linewidth of the TE$_{10}$ mode strongly narrows and the mode eventually vanishes at $k = 0$. Linewidth narrowing conjunct with the anomalous dispersion of the mode testify the presence of the BIC state predicted by the numerical calculations. We note that the BIC is observed even though a residual absorption of the MoSe$_2$ layer revealed by the ellipsometry measurements (see Fig. \ref{fig:techn}c). This indicates that the internal losses do not preclude the occurrence of the BIC, as further confirmed by simulations (see Fig.~S4). The 3D tomography measurements of the optical modes (see Fig.~S5 and Video 1 in the Supporting Information) show that the mode hosting the BIC exhibits a saddle-like dispersion. The BIC resides on a minimum of the dispersion along $k_{y}$ direction and on a maximum along $k_{x}$ direction.

In the reflection spectra of the TM-polarised light, we observe no optical modes in the considered spectral range, in agreement with the theoretical predictions. In consequence, lines indicating the cut-off of the first diffraction order propagation in the saphire substrate
(see also Supporting Information) are clearly visible, both in experiment and calculation results. 
The wavelength of the crossing of there lines (880~nm) corresponds to the product of the grating period and the refractive index of the sapphire ($n_{Al_2O_3}(\lambda=880\,\mathrm{nm})=1.76$). The very weak diffraction-limit lines exist also for the TE-polarized light, but they are barely visible due to the presence of the actual optical mode.

\begin{figure}[htp]
    \includegraphics[width=0.8\textwidth]{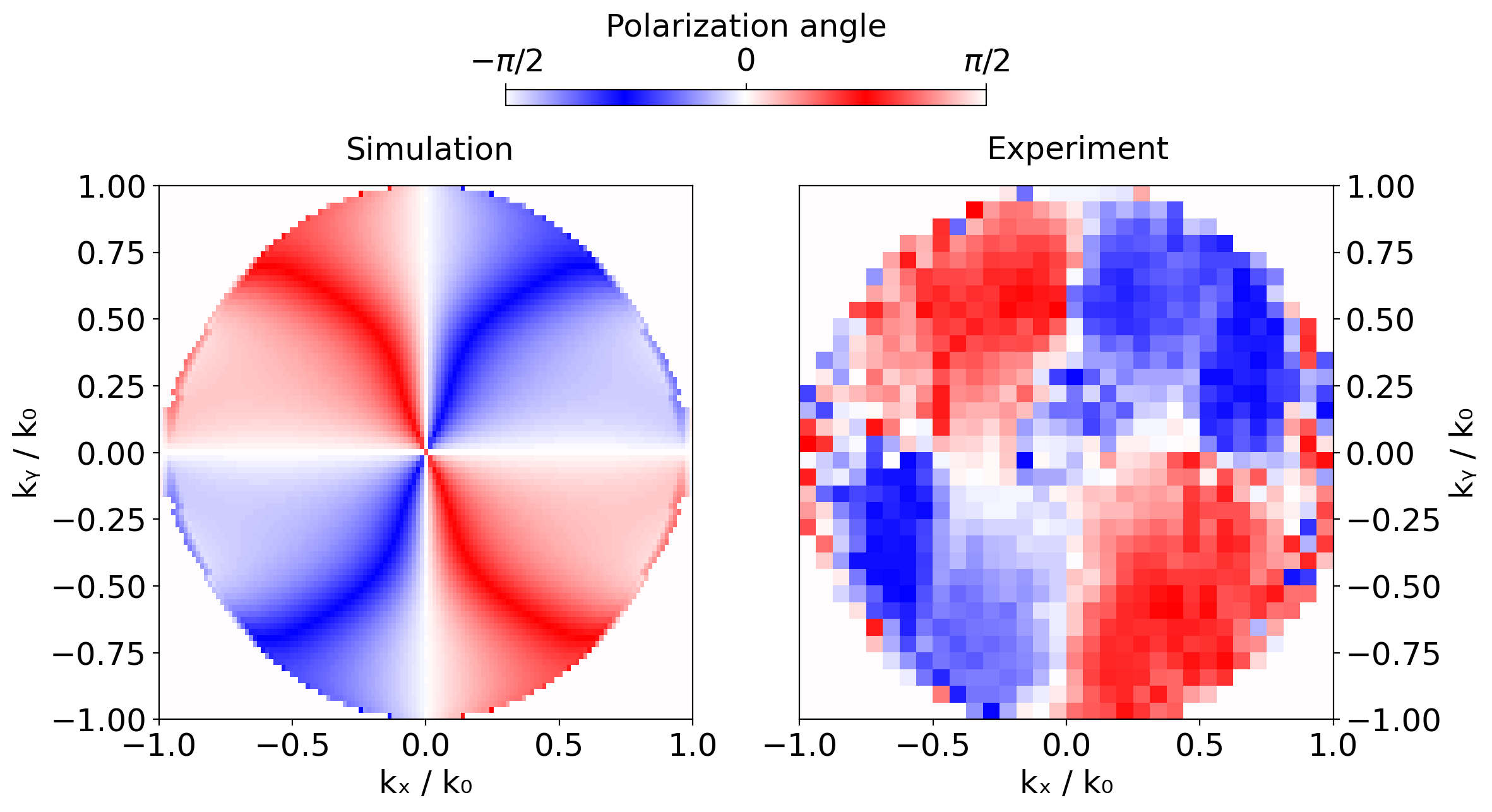}
    
    \caption{Polarization vortex of the anti-symmetric TE$_{10}$ mode around the BIC hosted in the minimum of its dispersion at $\lambda=1100\,\mathrm{nm}$ a) obtained numerically b) determined from the angle-resolved reflectivity experiment. The BIC acts as the vortex center for the polarization vector of the reflected light as revealed by the rotation of the direction of the polarization vector represented by the $\phi$ angle plotted as a function of perpendicular components of the wave vector $k_x$ and $k_y$ (relative to the wave vector length $k_0$).}
    \label{fig:vortex}
\end{figure}

To confirm the presence of the BIC, which is a vortex center in the polarization patterns of far-field radiation, characterized by conserved and quantized topological charges defined by the winding number of the polarization vectors \cite{Nguyen:Vortex,Zhen:PRL2013}. 
Figure~\ref{fig:vortex} shows a polarization vortex of the anti-symmetric TE$_{10}$ mode obtained numerically and by the experiment. Here, the horizontal and vertical polarizations correspond to the TE and TM polarizations, respectively.  BIC indeed acts as the vortex center for the polarization vector of the reflected light, as revealed by the rotation of the direction of the polarization vector represented by the $\phi$ angle plotted as a function of perpendicular components of the wave vector $k_x$ and $k_y$ (relative to the wave vector length $k_0$).
The existence of the vortex provides the final confirmation for the presence of a BIC in the studied structures.\cite{Zhen:PRL2013}

\textbf{Nonlinear investigations}
Finally, to demonstrate the low optical energy dissipation of the BIC state formed in an ultra-thin MoSe$_2$ grating and its ability to enhance light-matter interactions, we employ the subwavelength grating to enhance the generation of higher-order light harmonics. 
We note first, that the magnitude of nonlinear effects of layered materials, including MoSe$_2$, depends strongly on the thickness of the film \cite{Khan2020}. In a TMD monolayer, inversion symmetry is absent due to the asymmetric positioning of the selenium atoms in the top and bottom layers relative to the central molybdenum atomic layer. With the increase in the number of the TMD layers, the atomic arrangement within each layer compensates for the asymmetry of adjacent layers and creates interlayer symmetry. This restores the material's centrosymmetry and suppresses nonlinear effects such as second harmonic generation (see also a discussion in Supporting Information). For that reason, our research focuses on third-order nonlinear effects, particularly third harmonic generation, which is capable of occurring in centrosymmetric media.

\begin{figure}
\centering
    \begin{subfigure}{0.4\textwidth}
        \subcaption{}
        \includegraphics[width=\textwidth]{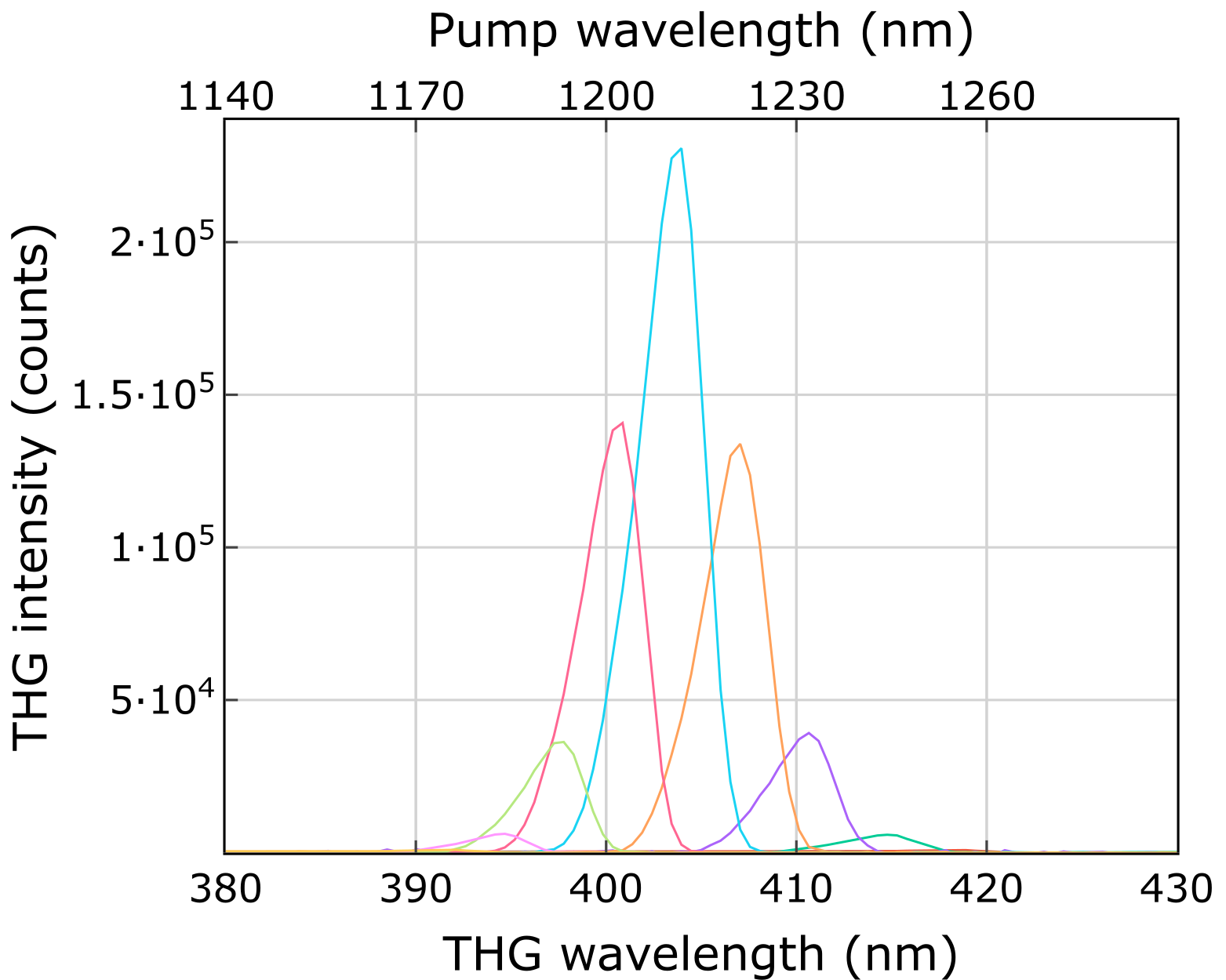}
    \end{subfigure}\hfill
    \begin{subfigure}{0.4\textwidth}
        \subcaption{}
        \includegraphics[width=\textwidth]{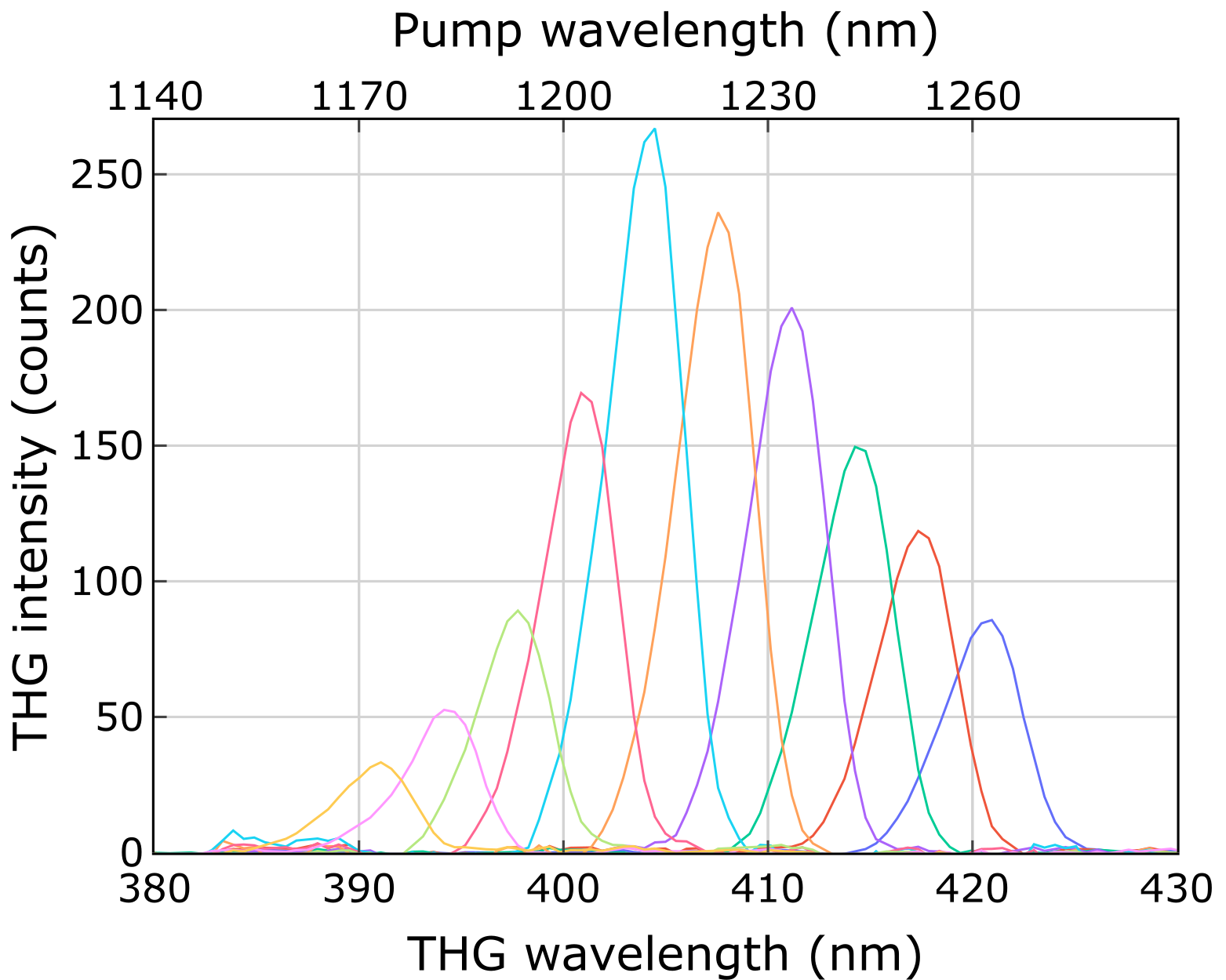}
    \end{subfigure}
    \begin{subfigure}{0.4\textwidth}
        \subcaption{}
        \includegraphics[width=\textwidth]{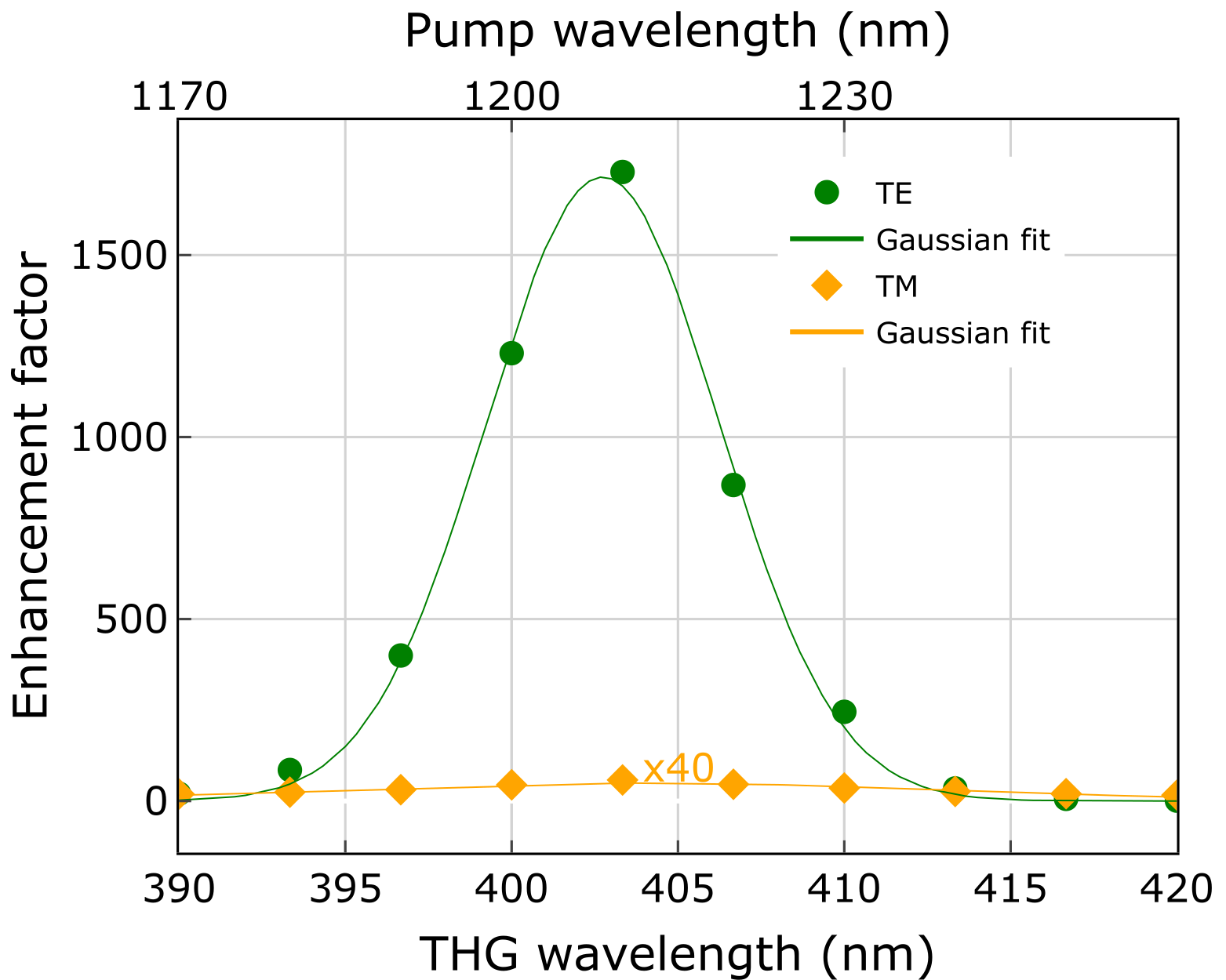}
    \end{subfigure}\hfill
    \begin{subfigure}[b]{0.4\textwidth}
        \subcaption{}
        \includegraphics[width=\textwidth]{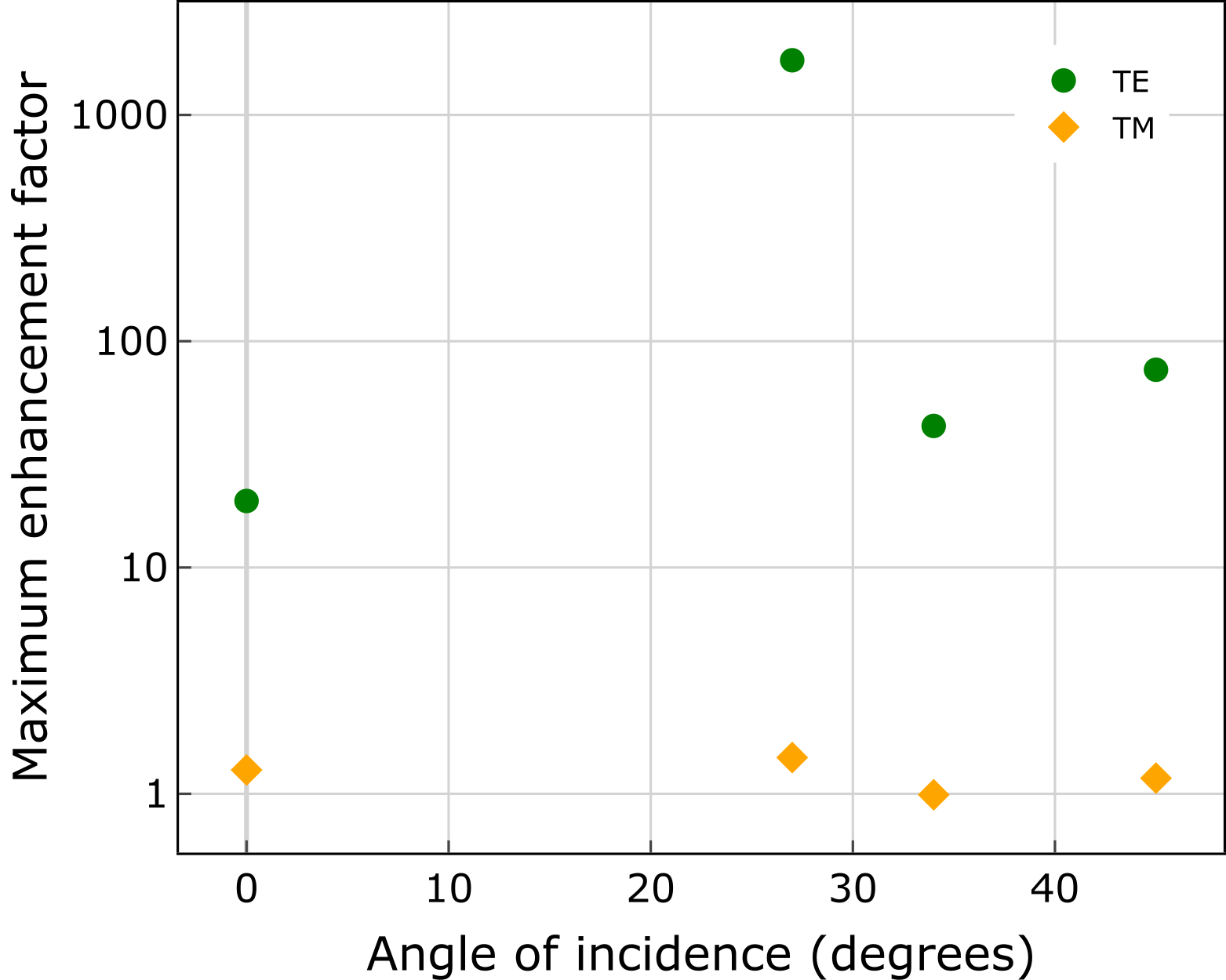}
    \end{subfigure}
    \caption{The nonlinear optical response of the MoSe$_2$ layer-based subwavelength grating. The intensity of THG generated under 27-degree excitation for a) TE and b) TM polarization. c) The wavelength dependence of the enhancement factor at a 27-degree angle for both polarizations. d) The angular dependence of the maximum observed enhancement factor.}
    \label{fig:THG_results}
\end{figure}

Although the structure exhibits the highest quality factor at normal incidence, the coupling efficiency of the pump beam into the structure under this angle is limited due to the radiationless nature of the BIC state. At higher incident angles, the coupling efficiency improves, which occurs, however, at the cost of reduced quality factor values. Aware of potential challenges associated with coupling light into the structure at normal incidence, the initial measurement is conducted at an incident angle of 27 degrees. At this angle, the structure's reflectivity in TE polarization reaches a minimum at the wavelength of approximately 1210 nm (see the dispersion of the optical mode in Fig.~\ref{fig:reflectivity_results}a), indicating the presence of a resonant optical mode within the SWG structure. The TH signal intensity as a function of the wavelength for TE polarization is shown in Fig.~\ref{fig:THG_results}a. As clearly seen, the wavelength corresponding to the strongest third-harmonic generation coincides with the wavelength at which the optical mode is observed in the grating. Interestingly, the TM polarized signal, although three orders of magnitude weaker than the TE polarized one, also exhibits its maximum intensity at the wavelength of the optical mode (see Fig.~\ref{fig:THG_results}b). We attribute this coincidence either to the imperfection of the fabricated structure or a slight misalignment of the grating with respect to the polarization of the incident light. In Figure \ref{fig:THG_results}c, we present the signal normalized relative to the reference TH signal intensity generated from a uniform, unstructured MoSe$_2$ layer of the same thickness. Although the exact BIC conditions are not achieved at a 27° angle, the TH signal intensity for TE polarization is increased by three orders of magnitude, with the enhancement factor reaching an impressive value of 1650.

The intensity of the TH generation as a function of the incidence angle is shown in Fig.~\ref{fig:THG_results}d. The lowest gain values, around 19, are observed for the 0-degree incidence angle, confirming that the coupling efficiency of the pump beam into the structure is limited at this angle. A reduction in the intensity of the THG is also found at angles of 34 and 45 degrees, for which the enhancement factor for the TE-polarized light is, respectively, around 40 and 70. In this case, however, we attribute a decrease in the efficiency of the TH generated signal to the reduction of the optical mode Q-factor values with the increase in in-plane photon momentum.
The detailed dependence of the TH signal intensity on the wavelength can be found in Figure S8 in the Supporting Information.

\section*{Conclusions and outlook}

We have exploited the exceptionally high refractive index of MoSe$_2$ to innovatively design and produce MoSe$_2$-based subwavelength gratings hosting BICs. The fabricated gratings,  with the thicknesses of tens of nanometers, are the thinnest structures hosting a BIC produced so far. The strong light confinement in the produced structures, conjunct with a high nonlinearity characteristic for the TMDs, has enabled a highly efficient third harmonic generation. We note that, while earlier studies reported BIC states and enhanced harmonic generation efficiency from TMD layers deposited on subwavelength gratings,\cite{Wang:Nanophotonics2024, Lee:NanoLett2025} our work demonstrates BIC and third-harmonic generation directly from a nanostructured layer of van der Waals material.

The demonstrated application of MBE for the production of a few inch-large, homogeneous structures ensures a scalability factor, crucial for possible industrial fabrication of ultrathin photonic structures. The advantage of scalability was not achievable until now, as prior demonstrations of TMD-based metasurfaces relied on exfoliated TMD flakes, which are limited in size to tens of $\mu$m and in thickness uniformity. The ease and simplicity of processing the structures confirm that other designs of photonic structures such as 2D metasurfaces based on TMD layers are feasible.

We envision that the future directions of the research will involve TMD heterostructures integrated with the TMD-based subwavelength gratings for boosting the emission intensity, as well as fabrication of the TMD-based subwavelength gratings with integrated single photon emitters, produced in a deterministic way within the grating, e.g., by e-beam irradiation.

\section*{Methods \label{Methods}}

\subsection{Numerical calculations}
For numerical calculations we employ Plane-Wave Admittance Method (PWAM) \cite{Dems2007PhD}. The PWAM allows us to determine optical modes and their field distributions. In addition, we use Plane-Wave Reflection Transformation Method (PWRTM) \cite{Dems2011OER} for calculations of reflection and transmission spectra of the structures. In both methods, we solve Maxwell's equations in a frequency domain by using a plane-wave expansion. We consider the refractive index anisotropy and the absorption of the materials by operating on tensorial complex permittivities. 

We calculated the quality factor (Q-factor) based on the values of the real $\lambda_{\mathrm{Re}}$ and imaginary part $\lambda_{\mathrm{Im}}$ of the determined mode wavelength. From the definition, the Q-factor is expressed as the ratio of the optical mode's central frequency to the mode's spectral width \cite{Chartier_Optics}. The central frequency is the real part of the complex frequency $\omega$ and the width of the resonance is equal to the double absolute value of the $\omega$ imaginary part $Q=\frac{\omega_{\mathrm{Re}}}{2 |\omega_{\mathrm{Im}}|}=\frac{|\lambda_{\mathrm{Im}}|}{2\lambda_{\mathrm{Re}}}$
\cite{Novotny2012}. We estimate Q-factor related to radiation losses $Q_{Rz}$ by calculating the Q-factor with an assumption of no absorption within the grating. Q-factor related to absorption $Q_{abs}$ is obtained in the result of comparing these calculations and calculations with absorption.

To calculate polarization vortices numerically, we first obtained the dispersion surface in $k_x$-$k_y$-$\lambda$ space and we simulated the near-field distributions of electric fields $E_x$ and $E_y$ for each point on the dispersion surface by using PWAM \cite{Dems2007PhD}. Then, we calculated the far-field distributions based on Huygens' principle \cite{EPK2014OQE} as $E'_x$ and $E'_y$, respectively. The polarization angle is calculated as $\phi_{th}=\arctan(\mathrm{real}(E'_x/E'_y))$.

\subsection{Sample growth}
Growth of the sample is performed in an MBE chamber delivered by SVT Associates \cite{Pacuski2020NL}. Surface roughness is controlled by Reflection High Energy Electron Diffraction (RHEED). Selenium with 99.9999\% purity (6N) is deposited from a standard low-temperature effusion cell and molybdenum with 99.995\% purity (4N5) is deposited using an e-beam with a rod (1/4" diameter, 35 mm length).

A growth procedure starts with heating a (0001) sapphire substrate with a 2-degree off-cut in an ultra-high vacuum to $700^{\circ}\mathrm{C}$. Next, the growth of MoSe$_2$ is performed at $300^{\circ}\mathrm{C}$ with a very small growth rate, about 1 monolayer (0.7 nm) per hour, optimized previously for the production of high optical-quality MoSe$_2$ monolayers. Every 20~h (14~nm of MoSe$_2$) the sample is additionally annealed at $700^{\circ}\mathrm{C}$ for 1~h to avoid the vertical growth the main challenge in the epitaxy of MoSe$_2$ layers and to keep its surface flat. The total thickness of MoSe$_2$ layer is 42~nm.

Additionally, after the growth, the samples are manually polished using silk in the environment of ethylene glycol for improving the surface quality. 

\subsection{Sample processing}

The MoSe$_2$ subwavelength gratings are fabricated through the reactive ion etching (RIE) technique. We use electron beam lithography to pattern the grating in 200 nm-thick CSAR 62 9\% positive resist spun for 60 s at 4000 rpm and baked on a hotplate at 150$^{\circ}$C for 15 min. The resist is additionally coated with a 40 nm-thick conducting polymer, Electra 92 (AR-PC 5090.02 by AllResist), spun at 4000 rpm for 60 s, and baked on a hotplate at 105$^{\circ}$C for 5~min. A polymer coating is used to dissipate electric charges effectively. The 100~\textmu m $\times$ 100~\textmu m grating pattern is exposed with an acceleration voltage of 30 keV using a beam current of 50~pA. The etching is done in an oxygen (O$_2$) and sulfur hexafluoride (SF$_6$) plasma in the Oxford Instrument PlasmaPro 100 machine, with the following parameters of the process: power of 300~W and pressure of 5 mTorr. The flow rate is set to 2~sccm for O$_2$, and to  20~sccm for SF$_6$. The etching lasts 20~seconds. After this process, a 42-nm-thick layer of MoSe$_2$ is completely removed from the patterned areas. The standard CSAR 62 remover AR 600-71 (by AllResist) is used to remove the resist leftovers. 

\subsection{Reflectivity measurements}
The schematic of the setup for the reflectivity measurements in k-space is presented in Fig.~S7a. 
A Xenon lamp (LDLS EQ-99X by Hamamatsu) serves as a spectrally broad source of light. The beam is collimated by a set of lenses, directed towards the sample by a non-polarizing (NP) 50:50 beam splitter (NIR BS), and then focused onto the sample surface using a near-infrared (NIR) microscope with a high NA of 0.7. The reflectivity signal from the sample is collected across all cone angles $\theta$, as shown in the inset to Fig.~S7d. The arrows depict the direction of the electric field of the light wave associated with TE and TM polarization. The reflected light is then directed toward Fourier imaging lenses for angular distribution measurements. The light passes then through a linear polarizer and a half-wave plate before being analyzed by a spectrometer. Fig.~S7e illustrates the beam alignment with the spectrometer slit, indicating the adjustments for accurate $k$-space imaging. The second lens in the Fourier imaging setup is motorized, allowing us to move the beam across spectrometer slit and measure different slices of the momentum space. The second lens in the Fourier imaging setup is motorized, enabling precise movement of the beam across the spectrometer slit to measure different slices of momentum space. Spectra are acquired using an iDus InGaAs CCD with a single line of 1024 pixels. To generate images like those in Fig.~\ref{fig:reflectivity_results}, the lens is moved vertically across the spectrometer slit (Fig.~S7d), capturing consecutive angles on the CCD. Each image showing experimental results in Fig.~\ref{fig:reflectivity_results} is composed of 128 individual spectra, which together form a single image.

To analyze the polarization vortex shown in Fig.~\ref{fig:vortex}, the entire $k$-space volume of the reflected light is sampled. This is accomplished by translating the beam across the slit both vertically and horizontally in a 32 by 32 grid pattern. This process yields successive slices of the $k$-space. By combining all these slices, a complete, spectrally resolved $k$-space volume is reconstructed, as illustrated in Fig.~S5. This dataset enables us to track the position of the mode in terms of energy and angular values, allowing for isolation or ``cutting'' of the mode from the $k$-space volume, as illustrated in Video 1 in the Supporting Information.

To demonstrate the presence the polarization vortex, we determine the Stokes parameters for the reflected light. This involves performing reflectivity measurements with the detection in four linear polarizations: horizontal ($I_{H}$), vertical ($I_{V}$), diagonal ($I_{D}$), and anti-diagonal ($I_{A}$). From these measurements, we can derive the Stokes components, defined as:
$$
S_{1} = \frac{I_{H} - I_{V}}{I_{H} + I_{V}}\, \mathrm{and}\, S_{2} = \frac{I_{D} - I_{A}}{I_{D} + I_{A}}.
$$
We then calculate the magnitude of the Stokes vector, $\rho$, given by:
$$
\rho = \sqrt{S_{1}^2 + S_{2}^2}.
$$
The polarization vector angle, $\phi$, can be determined by the relationship:
$$
\sin(2\phi) = \frac{S_{2}}{\rho}\, \mathrm{or}\, \cos(2\phi) = \frac{S_{1}}{\rho},
$$
leading to the calculation:
$$
\phi = \frac{1}{2} \arcsin\left(\frac{S_{2}}{\rho}\right).
$$

\subsection{Ellipsometry measurements}
The optical quality of the as-grown MoSe$_2$ layer is examined using variable angle spectroscopic ellipsometry (SE). Our investigations are conducted with an RC2 ellipsometer (manufactured by J.A. Woollam Co.) in the spectral range from 193 nm to 1690 nm. Considering that both the sample and the substrate may exhibit birefringence, leading to possible cross-polarization and depolarization effects, we measure all 16 elements of the Mueller matrix (MM), both in reflection (angle of incidence from 55° to 70° by 5°) and transmission (angle of incidence from 0° to 40° by 5°). The analysis of the SE data is complemented by transmission intensity measurements taken at angles from 0° to 40°. This combined approach reduces the possible correlation between sample thickness and optical constants, ensuring an unequivocal data description. The substrate bottom surface is roughened to eliminate backside reflections in reflection measurements. The MoSe$_2$ dielectric permittivity model assumes that the MoSe$_2$ layer is uniaxial, with several Tauc-Lorentz oscillators in the in-plane direction.

\subsection{Third harmonic generation}
The nonlinear measurements are performed using two distinct experimental reflection setups (see Figure S7). In the first configuration (Fig. S7b), two microscope objectives are used: one to illuminate the sample with a near-infrared beam (Mitutoyo MY10X-823 with NA = 0.26) and the other to collect the emerging third harmonic signal (Mitutoyo MY20X-804 with NA = 0.42) and direct it toward a spectrometer (Isoplane 320 from Teledyne Princeton Instruments). This arrangement covers angles from 27° to nearly 90° and is more efficient in terms of signal-to-noise ratio. For the 0° configuration, due to constraints imposed by the system’s geometry, a single objective (Olympus UPLFLN with NA = 0.3) fulfills both illumination and collection functions (Configuration Fig. S7c). An amplified femtosecond laser (PHAROS-SP-HP, Light Conversion) together with an optical parametric amplifier (ORPHEUS-HP, Light Conversion) is used to generate light pulses in the wavelength range of 1050–1350 nm. The peak power focused on the sample surface is on the order of 100 GW/cm², and the sample position is adjusted for every illumination wavelength to maximize the third harmonic (TH) signal. The THG measurement for a given wavelength is repeated several times in different areas of the grating. The reference signal is collected from the same sample, but from an area without the grating.

\section*{Supporting Information}

\begin{itemize}
\item Refractive index of the sapphire substrate; Additional results of numerical calculations; Reflectivity and transmission of the epitaxial layers; Optical microscope image of the sample surface; 3D tomography of the reflectivity of the gratings; Schemes of the experimental setups; Additional results of the nonlinear investigation.
\item Video presenting 3D tomography of the reflectivity of the gratings.
\end{itemize}

\section*{Acknowledgements}
This work was supported by the Polish National Science Center within the projects OPUS 2020/39/B/ST7/03502 and 2021/41/B/ST3/04183. The University of Warsaw supported this work within the ``Excellence Initiative -- Research University'' (IDUB) New Ideas 3A in Priority Research Area II grant 501-D111-20-2004310 ''Ultracienkie siatki podfalowe na bazie dichalkogenków''.

The work was also supported by the European Union through the ERC-ADVANCED grant TERAPLASM (No. 101053716). Views and opinions expressed are, however, those of the author(s) only and do not necessarily reflect those of the European Union or the European Research Council Executive Agency. Neither the European Union nor the granting authority can be held responsible for them. We also acknowledge the support of the "Center for Terahertz Research and Applications (CENTERA2)" project (FENG.02.01-IP.05-T004/23) carried out within the "International Research Agendas" program of the Foundation for Polish Science co-financed by the European Union under European Funds for a Smart Economy Programme.

\bibliography{MoSe2}

\begin{thebibliography}{76}%
\makeatletter
\providecommand \@ifxundefined [1]{%
 \@ifx{#1\undefined}
}%
\providecommand \@ifnum [1]{%
 \ifnum #1\expandafter \@firstoftwo
 \else \expandafter \@secondoftwo
 \fi
}%
\providecommand \@ifx [1]{%
 \ifx #1\expandafter \@firstoftwo
 \else \expandafter \@secondoftwo
 \fi
}%
\providecommand \natexlab [1]{#1}%
\providecommand \enquote  [1]{``#1''}%
\providecommand \bibnamefont  [1]{#1}%
\providecommand \bibfnamefont [1]{#1}%
\providecommand \citenamefont [1]{#1}%
\providecommand \href@noop [0]{\@secondoftwo}%
\providecommand \href [0]{\begingroup \@sanitize@url \@href}%
\providecommand \@href[1]{\@@startlink{#1}\@@href}%
\providecommand \@@href[1]{\endgroup#1\@@endlink}%
\providecommand \@sanitize@url [0]{\catcode `\\12\catcode `\$12\catcode
  `\&12\catcode `\#12\catcode `\^12\catcode `\_12\catcode `\%12\relax}%
\providecommand \@@startlink[1]{}%
\providecommand \@@endlink[0]{}%
\providecommand \url  [0]{\begingroup\@sanitize@url \@url }%
\providecommand \@url [1]{\endgroup\@href {#1}{\urlprefix }}%
\providecommand \urlprefix  [0]{URL }%
\providecommand \Eprint [0]{\href }%
\providecommand \doibase [0]{https://doi.org/}%
\providecommand \selectlanguage [0]{\@gobble}%
\providecommand \bibinfo  [0]{\@secondoftwo}%
\providecommand \bibfield  [0]{\@secondoftwo}%
\providecommand \translation [1]{[#1]}%
\providecommand \BibitemOpen [0]{}%
\providecommand \bibitemStop [0]{}%
\providecommand \bibitemNoStop [0]{.\EOS\space}%
\providecommand \EOS [0]{\spacefactor3000\relax}%
\providecommand \BibitemShut  [1]{\csname bibitem#1\endcsname}%
\let\auto@bib@innerbib\@empty
\bibitem [{\citenamefont {Marinica}\ \emph {et~al.}(2008)\citenamefont
  {Marinica}, \citenamefont {Borisov},\ and\ \citenamefont
  {Shabanov}}]{Marinica2008PRL}%
  \BibitemOpen
  \bibfield  {author} {\bibinfo {author} {\bibfnamefont {D.~C.}\ \bibnamefont
  {Marinica}}, \bibinfo {author} {\bibfnamefont {A.~G.}\ \bibnamefont
  {Borisov}},\ and\ \bibinfo {author} {\bibfnamefont {S.~V.}\ \bibnamefont
  {Shabanov}},\ }\bibfield  {title} {\bibinfo {title} {Bound states in the
  continuum in photonics},\ }\href
  {https://doi.org/10.1103/PhysRevLett.100.183902} {\bibfield  {journal}
  {\bibinfo  {journal} {Physical Review Letters}\ }\textbf {\bibinfo {volume}
  {100}},\ \bibinfo {pages} {1} (\bibinfo {year} {2008})}\BibitemShut {NoStop}%
\bibitem [{\citenamefont {Plotnik}\ \emph {et~al.}(2011)\citenamefont
  {Plotnik}, \citenamefont {Peleg}, \citenamefont {Dreisow}, \citenamefont
  {Heinrich}, \citenamefont {Nolte}, \citenamefont {Szameit},\ and\
  \citenamefont {Segev}}]{Plotnik2011PRL}%
  \BibitemOpen
  \bibfield  {author} {\bibinfo {author} {\bibfnamefont {Y.}~\bibnamefont
  {Plotnik}}, \bibinfo {author} {\bibfnamefont {O.}~\bibnamefont {Peleg}},
  \bibinfo {author} {\bibfnamefont {F.}~\bibnamefont {Dreisow}}, \bibinfo
  {author} {\bibfnamefont {M.}~\bibnamefont {Heinrich}}, \bibinfo {author}
  {\bibfnamefont {S.}~\bibnamefont {Nolte}}, \bibinfo {author} {\bibfnamefont
  {A.}~\bibnamefont {Szameit}},\ and\ \bibinfo {author} {\bibfnamefont
  {M.}~\bibnamefont {Segev}},\ }\bibfield  {title} {\bibinfo {title}
  {Experimental observation of optical bound states in the continuum},\ }\href
  {https://doi.org/10.1103/PhysRevLett.107.183901} {\bibfield  {journal}
  {\bibinfo  {journal} {Physical Review Letters}\ }\textbf {\bibinfo {volume}
  {107}},\ \bibinfo {pages} {28} (\bibinfo {year} {2011})}\BibitemShut
  {NoStop}%
\bibitem [{\citenamefont {Robbiano}\ \emph {et~al.}(2018)\citenamefont
  {Robbiano}, \citenamefont {Paternò}, \citenamefont {Mattina}, \citenamefont
  {Motti}, \citenamefont {Lanzani}, \citenamefont {Scotognella},\ and\
  \citenamefont {Barillaro}}]{Robbiano2018AN}%
  \BibitemOpen
  \bibfield  {author} {\bibinfo {author} {\bibfnamefont {V.}~\bibnamefont
  {Robbiano}}, \bibinfo {author} {\bibfnamefont {G.~M.}\ \bibnamefont
  {Paternò}}, \bibinfo {author} {\bibfnamefont {A.~A.~L.}\ \bibnamefont
  {Mattina}}, \bibinfo {author} {\bibfnamefont {S.~G.}\ \bibnamefont {Motti}},
  \bibinfo {author} {\bibfnamefont {G.}~\bibnamefont {Lanzani}}, \bibinfo
  {author} {\bibfnamefont {F.}~\bibnamefont {Scotognella}},\ and\ \bibinfo
  {author} {\bibfnamefont {G.}~\bibnamefont {Barillaro}},\ }\bibfield  {title}
  {\bibinfo {title} {Room-temperature low-threshold lasing from monolithically
  integrated nanostructured porous silicon hybrid microcavities},\ }\href
  {https://doi.org/10.1021/acsnano.8b00875} {\bibfield  {journal} {\bibinfo
  {journal} {ACS Nano}\ }\textbf {\bibinfo {volume} {12}},\ \bibinfo {pages}
  {4536} (\bibinfo {year} {2018})}\BibitemShut {NoStop}%
\bibitem [{\citenamefont {Wang}\ \emph {et~al.}(2019)\citenamefont {Wang},
  \citenamefont {He}, \citenamefont {Chung}, \citenamefont {Hu}, \citenamefont
  {Yu}, \citenamefont {Chen}, \citenamefont {Ding}, \citenamefont {Chen},
  \citenamefont {Qin}, \citenamefont {Yang}, \citenamefont {Liu}, \citenamefont
  {Duan}, \citenamefont {Li}, \citenamefont {Gerhardt}, \citenamefont
  {Winkler}, \citenamefont {Jurkat}, \citenamefont {Wang}, \citenamefont
  {Gregersen}, \citenamefont {Huo}, \citenamefont {Dai}, \citenamefont {Yu},
  \citenamefont {Höfling}, \citenamefont {Lu},\ and\ \citenamefont
  {Pan}}]{Wang2019NPh}%
  \BibitemOpen
  \bibfield  {author} {\bibinfo {author} {\bibfnamefont {H.}~\bibnamefont
  {Wang}}, \bibinfo {author} {\bibfnamefont {Y.~M.}\ \bibnamefont {He}},
  \bibinfo {author} {\bibfnamefont {T.~H.}\ \bibnamefont {Chung}}, \bibinfo
  {author} {\bibfnamefont {H.}~\bibnamefont {Hu}}, \bibinfo {author}
  {\bibfnamefont {Y.}~\bibnamefont {Yu}}, \bibinfo {author} {\bibfnamefont
  {S.}~\bibnamefont {Chen}}, \bibinfo {author} {\bibfnamefont {X.}~\bibnamefont
  {Ding}}, \bibinfo {author} {\bibfnamefont {M.~C.}\ \bibnamefont {Chen}},
  \bibinfo {author} {\bibfnamefont {J.}~\bibnamefont {Qin}}, \bibinfo {author}
  {\bibfnamefont {X.}~\bibnamefont {Yang}}, \bibinfo {author} {\bibfnamefont
  {R.~Z.}\ \bibnamefont {Liu}}, \bibinfo {author} {\bibfnamefont {Z.~C.}\
  \bibnamefont {Duan}}, \bibinfo {author} {\bibfnamefont {J.~P.}\ \bibnamefont
  {Li}}, \bibinfo {author} {\bibfnamefont {S.}~\bibnamefont {Gerhardt}},
  \bibinfo {author} {\bibfnamefont {K.}~\bibnamefont {Winkler}}, \bibinfo
  {author} {\bibfnamefont {J.}~\bibnamefont {Jurkat}}, \bibinfo {author}
  {\bibfnamefont {L.~J.}\ \bibnamefont {Wang}}, \bibinfo {author}
  {\bibfnamefont {N.}~\bibnamefont {Gregersen}}, \bibinfo {author}
  {\bibfnamefont {Y.~H.}\ \bibnamefont {Huo}}, \bibinfo {author} {\bibfnamefont
  {Q.}~\bibnamefont {Dai}}, \bibinfo {author} {\bibfnamefont {S.}~\bibnamefont
  {Yu}}, \bibinfo {author} {\bibfnamefont {S.}~\bibnamefont {Höfling}},
  \bibinfo {author} {\bibfnamefont {C.~Y.}\ \bibnamefont {Lu}},\ and\ \bibinfo
  {author} {\bibfnamefont {J.~W.}\ \bibnamefont {Pan}},\ }\bibfield  {title}
  {\bibinfo {title} {Towards optimal single-photon sources from polarized
  microcavities},\ }\href {https://doi.org/10.1038/s41566-019-0494-3}
  {\bibfield  {journal} {\bibinfo  {journal} {Nature Photonics}\ }\textbf
  {\bibinfo {volume} {13}},\ \bibinfo {pages} {770} (\bibinfo {year}
  {2019})}\BibitemShut {NoStop}%
\bibitem [{\citenamefont {Hakala}\ \emph {et~al.}(2018)\citenamefont {Hakala},
  \citenamefont {Moilanen}, \citenamefont {Väkeväinen}, \citenamefont {Guo},
  \citenamefont {Martikainen}, \citenamefont {Daskalakis}, \citenamefont
  {Rekola}, \citenamefont {Julku},\ and\ \citenamefont
  {Törmä}}]{Hakala2018NP}%
  \BibitemOpen
  \bibfield  {author} {\bibinfo {author} {\bibfnamefont {T.~K.}\ \bibnamefont
  {Hakala}}, \bibinfo {author} {\bibfnamefont {A.~J.}\ \bibnamefont
  {Moilanen}}, \bibinfo {author} {\bibfnamefont {A.~I.}\ \bibnamefont
  {Väkeväinen}}, \bibinfo {author} {\bibfnamefont {R.}~\bibnamefont {Guo}},
  \bibinfo {author} {\bibfnamefont {J.~P.}\ \bibnamefont {Martikainen}},
  \bibinfo {author} {\bibfnamefont {K.~S.}\ \bibnamefont {Daskalakis}},
  \bibinfo {author} {\bibfnamefont {H.~T.}\ \bibnamefont {Rekola}}, \bibinfo
  {author} {\bibfnamefont {A.}~\bibnamefont {Julku}},\ and\ \bibinfo {author}
  {\bibfnamefont {P.}~\bibnamefont {Törmä}},\ }\bibfield  {title} {\bibinfo
  {title} {Bose-einstein condensation in a plasmonic lattice},\ }\href
  {https://doi.org/10.1038/s41567-018-0109-9} {\bibfield  {journal} {\bibinfo
  {journal} {Nature Physics}\ }\textbf {\bibinfo {volume} {14}},\ \bibinfo
  {pages} {739} (\bibinfo {year} {2018})}\BibitemShut {NoStop}%
\bibitem [{\citenamefont {He}\ \emph {et~al.}(2020)\citenamefont {He},
  \citenamefont {Harris}, \citenamefont {Baker}, \citenamefont {Sawadsky},
  \citenamefont {Sfendla}, \citenamefont {Sachkou}, \citenamefont {Forstner},\
  and\ \citenamefont {Bowen}}]{He2020NP}%
  \BibitemOpen
  \bibfield  {author} {\bibinfo {author} {\bibfnamefont {X.}~\bibnamefont
  {He}}, \bibinfo {author} {\bibfnamefont {G.~I.}\ \bibnamefont {Harris}},
  \bibinfo {author} {\bibfnamefont {C.~G.}\ \bibnamefont {Baker}}, \bibinfo
  {author} {\bibfnamefont {A.}~\bibnamefont {Sawadsky}}, \bibinfo {author}
  {\bibfnamefont {Y.~L.}\ \bibnamefont {Sfendla}}, \bibinfo {author}
  {\bibfnamefont {Y.~P.}\ \bibnamefont {Sachkou}}, \bibinfo {author}
  {\bibfnamefont {S.}~\bibnamefont {Forstner}},\ and\ \bibinfo {author}
  {\bibfnamefont {W.~P.}\ \bibnamefont {Bowen}},\ }\bibfield  {title} {\bibinfo
  {title} {Strong optical coupling through superfluid brillouin lasing},\
  }\href {https://doi.org/10.1038/s41567-020-0785-0} {\bibfield  {journal}
  {\bibinfo  {journal} {Nature Physics}\ }\textbf {\bibinfo {volume} {16}},\
  \bibinfo {pages} {417} (\bibinfo {year} {2020})}\BibitemShut {NoStop}%
\bibitem [{\citenamefont {Głowadzka}\ \emph {et~al.}(2021)\citenamefont
  {Głowadzka}, \citenamefont {Wasiak},\ and\ \citenamefont
  {Czyszanowski}}]{Glowadzka2021NPh}%
  \BibitemOpen
  \bibfield  {author} {\bibinfo {author} {\bibfnamefont {W.}~\bibnamefont
  {Głowadzka}}, \bibinfo {author} {\bibfnamefont {M.}~\bibnamefont {Wasiak}},\
  and\ \bibinfo {author} {\bibfnamefont {T.}~\bibnamefont {Czyszanowski}},\
  }\bibfield  {title} {\bibinfo {title} {True- and quasi-bound states in the
  continuum in one-dimensional gratings with broken up-down mirror symmetry:
  Bound states in gratings with broken symmetry},\ }\href
  {https://doi.org/10.1515/nanoph-2021-0319} {\bibfield  {journal} {\bibinfo
  {journal} {Nanophotonics}\ }\textbf {\bibinfo {volume} {10}},\ \bibinfo
  {pages} {3979} (\bibinfo {year} {2021})}\BibitemShut {NoStop}%
\bibitem [{\citenamefont {Adachi}(1989)}]{Adachi1989JAP}%
  \BibitemOpen
  \bibfield  {author} {\bibinfo {author} {\bibfnamefont {S.}~\bibnamefont
  {Adachi}},\ }\bibfield  {title} {\bibinfo {title} {Optical dispersion
  relations for gap, gaas, gasb, inp, inas, insb, al$_x$ga$_{1-x}$as, and
  in$_{1-x}$ga$_x$as$_y$p$_{1-y}$},\ }\href {https://doi.org/10.1063/1.343580}
  {\bibfield  {journal} {\bibinfo  {journal} {Journal of Applied Physics}\
  }\textbf {\bibinfo {volume} {66}},\ \bibinfo {pages} {6030} (\bibinfo {year}
  {1989})}\BibitemShut {NoStop}%
\bibitem [{\citenamefont {Rakić}\ and\ \citenamefont
  {Majewski}(1996)}]{Rakic1996JAP}%
  \BibitemOpen
  \bibfield  {author} {\bibinfo {author} {\bibfnamefont {A.~D.}\ \bibnamefont
  {Rakić}}\ and\ \bibinfo {author} {\bibfnamefont {M.~L.}\ \bibnamefont
  {Majewski}},\ }\href {https://doi.org/10.1063/1.363586} {\bibinfo {title}
  {Modeling the optical dielectric function of {GaAs} and {AlAs}: Extension of
  {Adachi}'s model}} (\bibinfo {year} {1996})\BibitemShut {NoStop}%
\bibitem [{\citenamefont {Barker}\ and\ \citenamefont
  {Ilegems}(1973)}]{Barker1973PRB}%
  \BibitemOpen
  \bibfield  {author} {\bibinfo {author} {\bibfnamefont {A.~S.}\ \bibnamefont
  {Barker}}\ and\ \bibinfo {author} {\bibfnamefont {M.}~\bibnamefont
  {Ilegems}},\ }\bibfield  {title} {\bibinfo {title} {Infrared lattice
  vibrations and free-electron dispersion in gan},\ }\href@noop {} {\bibfield
  {journal} {\bibinfo  {journal} {Physical Review B}\ }\textbf {\bibinfo
  {volume} {7}},\ \bibinfo {pages} {743} (\bibinfo {year} {1973})}\BibitemShut
  {NoStop}%
\bibitem [{\citenamefont {Beliaev}\ \emph {et~al.}(2021)\citenamefont
  {Beliaev}, \citenamefont {Shkondin}, \citenamefont {Lavrinenko},\ and\
  \citenamefont {Takayama}}]{Beliaev2021JVST}%
  \BibitemOpen
  \bibfield  {author} {\bibinfo {author} {\bibfnamefont {L.~Y.}\ \bibnamefont
  {Beliaev}}, \bibinfo {author} {\bibfnamefont {E.}~\bibnamefont {Shkondin}},
  \bibinfo {author} {\bibfnamefont {A.~V.}\ \bibnamefont {Lavrinenko}},\ and\
  \bibinfo {author} {\bibfnamefont {O.}~\bibnamefont {Takayama}},\ }\bibfield
  {title} {\bibinfo {title} {Thickness-dependent optical properties of aluminum
  nitride films for mid-infrared wavelengths},\ }\bibfield  {journal} {\bibinfo
   {journal} {Journal of Vacuum Science \& Technology A: Vacuum, Surfaces, and
  Films}\ }\textbf {\bibinfo {volume} {39}},\ \href
  {https://doi.org/10.1116/6.0000884} {10.1116/6.0000884} (\bibinfo {year}
  {2021})\BibitemShut {NoStop}%
\bibitem [{\citenamefont {Ferrini}\ \emph {et~al.}(1998)\citenamefont
  {Ferrini}, \citenamefont {Patrini},\ and\ \citenamefont
  {Franchi}}]{Ferrini1998JAP}%
  \BibitemOpen
  \bibfield  {author} {\bibinfo {author} {\bibfnamefont {R.}~\bibnamefont
  {Ferrini}}, \bibinfo {author} {\bibfnamefont {M.}~\bibnamefont {Patrini}},\
  and\ \bibinfo {author} {\bibfnamefont {S.}~\bibnamefont {Franchi}},\
  }\bibfield  {title} {\bibinfo {title} {Optical functions from 0.02 to 6 ev of
  al$_x$ga$_{1-x}$sb/gasb epitaxial layers},\ }\href
  {https://doi.org/https://doi.org/10.1063/1.368677} {\bibfield  {journal}
  {\bibinfo  {journal} {Journal of Applied Physics}\ }\textbf {\bibinfo
  {volume} {84}},\ \bibinfo {pages} {4517} (\bibinfo {year}
  {1998})}\BibitemShut {NoStop}%
\bibitem [{\citenamefont {Schinke}\ \emph {et~al.}(2015)\citenamefont
  {Schinke}, \citenamefont {Peest}, \citenamefont {Schmidt}, \citenamefont
  {Brendel}, \citenamefont {Bothe}, \citenamefont {Vogt}, \citenamefont
  {Kröger}, \citenamefont {Winter}, \citenamefont {Schirmacher}, \citenamefont
  {Lim}, \citenamefont {Nguyen},\ and\ \citenamefont
  {Macdonald}}]{Schinke2015AIPA}%
  \BibitemOpen
  \bibfield  {author} {\bibinfo {author} {\bibfnamefont {C.}~\bibnamefont
  {Schinke}}, \bibinfo {author} {\bibfnamefont {P.~C.}\ \bibnamefont {Peest}},
  \bibinfo {author} {\bibfnamefont {J.}~\bibnamefont {Schmidt}}, \bibinfo
  {author} {\bibfnamefont {R.}~\bibnamefont {Brendel}}, \bibinfo {author}
  {\bibfnamefont {K.}~\bibnamefont {Bothe}}, \bibinfo {author} {\bibfnamefont
  {M.~R.}\ \bibnamefont {Vogt}}, \bibinfo {author} {\bibfnamefont
  {I.}~\bibnamefont {Kröger}}, \bibinfo {author} {\bibfnamefont
  {S.}~\bibnamefont {Winter}}, \bibinfo {author} {\bibfnamefont
  {A.}~\bibnamefont {Schirmacher}}, \bibinfo {author} {\bibfnamefont
  {S.}~\bibnamefont {Lim}}, \bibinfo {author} {\bibfnamefont {H.~T.}\
  \bibnamefont {Nguyen}},\ and\ \bibinfo {author} {\bibfnamefont
  {D.}~\bibnamefont {Macdonald}},\ }\bibfield  {title} {\bibinfo {title}
  {Uncertainty analysis for the coefficient of band-to-band absorption of
  crystalline silicon},\ }\bibfield  {journal} {\bibinfo  {journal} {AIP
  Advances}\ }\textbf {\bibinfo {volume} {5}},\ \href
  {https://doi.org/10.1063/1.4923379} {10.1063/1.4923379} (\bibinfo {year}
  {2015})\BibitemShut {NoStop}%
\bibitem [{\citenamefont {Amotchkina}\ \emph {et~al.}(2020)\citenamefont
  {Amotchkina}, \citenamefont {Trubetskov}, \citenamefont {Hahner},\ and\
  \citenamefont {Pervak}}]{Amotchkina2020AO}%
  \BibitemOpen
  \bibfield  {author} {\bibinfo {author} {\bibfnamefont {T.}~\bibnamefont
  {Amotchkina}}, \bibinfo {author} {\bibfnamefont {M.}~\bibnamefont
  {Trubetskov}}, \bibinfo {author} {\bibfnamefont {D.}~\bibnamefont {Hahner}},\
  and\ \bibinfo {author} {\bibfnamefont {V.}~\bibnamefont {Pervak}},\
  }\bibfield  {title} {\bibinfo {title} {Characterization of e-beam evaporated
  ge, ybf$_3$, zns, and laf$_3$ thin films for laser-oriented coatings},\
  }\href {https://doi.org/10.1364/AO.59.000A40} {\bibfield  {journal} {\bibinfo
   {journal} {Appl. Opt.}\ }\textbf {\bibinfo {volume} {59}},\ \bibinfo {pages}
  {A40} (\bibinfo {year} {2020})}\BibitemShut {NoStop}%
\bibitem [{\citenamefont {Larruquert}\ \emph {et~al.}(2011)\citenamefont
  {Larruquert}, \citenamefont {Pérez-Marín}, \citenamefont {García-Cortés},
  \citenamefont {{Rodríguez-De Marcos}}, \citenamefont {Aznárez},\ and\
  \citenamefont {Méndez}}]{Larruquert2011JOSAA}%
  \BibitemOpen
  \bibfield  {author} {\bibinfo {author} {\bibfnamefont {J.~I.}\ \bibnamefont
  {Larruquert}}, \bibinfo {author} {\bibfnamefont {A.~P.}\ \bibnamefont
  {Pérez-Marín}}, \bibinfo {author} {\bibfnamefont {S.}~\bibnamefont
  {García-Cortés}}, \bibinfo {author} {\bibfnamefont {L.}~\bibnamefont
  {{Rodríguez-De Marcos}}}, \bibinfo {author} {\bibfnamefont {J.~A.}\
  \bibnamefont {Aznárez}},\ and\ \bibinfo {author} {\bibfnamefont {J.~A.}\
  \bibnamefont {Méndez}},\ }\bibfield  {title} {\bibinfo {title}
  {Self-consistent optical constants of sic thin films},\ }\href@noop {}
  {\bibfield  {journal} {\bibinfo  {journal} {Journal of the Optical Society of
  America A}\ }\textbf {\bibinfo {volume} {28}},\ \bibinfo {pages} {2340}
  (\bibinfo {year} {2011})}\BibitemShut {NoStop}%
\bibitem [{\citenamefont {Pflüger}\ \emph {et~al.}(1984)\citenamefont
  {Pflüger}, \citenamefont {Fink}, \citenamefont {Weber}, \citenamefont
  {Bohnen},\ and\ \citenamefont {Crecelius}}]{Pfluger1984PRB}%
  \BibitemOpen
  \bibfield  {author} {\bibinfo {author} {\bibfnamefont {J.}~\bibnamefont
  {Pflüger}}, \bibinfo {author} {\bibfnamefont {J.}~\bibnamefont {Fink}},
  \bibinfo {author} {\bibfnamefont {W.}~\bibnamefont {Weber}}, \bibinfo
  {author} {\bibfnamefont {K.-P.}\ \bibnamefont {Bohnen}},\ and\ \bibinfo
  {author} {\bibfnamefont {G.}~\bibnamefont {Crecelius}},\ }\bibfield  {title}
  {\bibinfo {title} {Dielectric properties of tic$_x$, tin$_x$, vc$_x$, and
  vn$_x$ from 1.5 to 40 ev determined by electron-energy-loss spectroscopy},\
  }\href@noop {} {\bibfield  {journal} {\bibinfo  {journal} {Physical Review
  B}\ }\textbf {\bibinfo {volume} {30}},\ \bibinfo {pages} {1155} (\bibinfo
  {year} {1984})}\BibitemShut {NoStop}%
\bibitem [{\citenamefont {Phillip}\ and\ \citenamefont
  {Taft}(1964)}]{Philip1964PR}%
  \BibitemOpen
  \bibfield  {author} {\bibinfo {author} {\bibfnamefont {H.~R.}\ \bibnamefont
  {Phillip}}\ and\ \bibinfo {author} {\bibfnamefont {E.~A.}\ \bibnamefont
  {Taft}},\ }\bibfield  {title} {\bibinfo {title} {Kramers-kronig analysis of
  reflectance data for diamond},\ }\href@noop {} {\bibfield  {journal}
  {\bibinfo  {journal} {Physical Rewiev}\ }\textbf {\bibinfo {volume} {136}},\
  \bibinfo {pages} {1445} (\bibinfo {year} {1964})}\BibitemShut {NoStop}%
\bibitem [{\citenamefont {Larruquert}\ \emph {et~al.}(2013)\citenamefont
  {Larruquert}, \citenamefont {{Rodríguez-de Marcos}}, \citenamefont
  {Méndez}, \citenamefont {Martin},\ and\ \citenamefont
  {Bendavid}}]{Larruquert2013OE}%
  \BibitemOpen
  \bibfield  {author} {\bibinfo {author} {\bibfnamefont {J.~I.}\ \bibnamefont
  {Larruquert}}, \bibinfo {author} {\bibfnamefont {L.~V.}\ \bibnamefont
  {{Rodríguez-de Marcos}}}, \bibinfo {author} {\bibfnamefont {J.~A.}\
  \bibnamefont {Méndez}}, \bibinfo {author} {\bibfnamefont {P.~J.}\
  \bibnamefont {Martin}},\ and\ \bibinfo {author} {\bibfnamefont
  {A.}~\bibnamefont {Bendavid}},\ }\bibfield  {title} {\bibinfo {title} {High
  reflectance ta-c coatings in the extreme ultraviolet},\ }\href
  {https://doi.org/10.1364/oe.21.027537} {\bibfield  {journal} {\bibinfo
  {journal} {Optics Express}\ }\textbf {\bibinfo {volume} {21}},\ \bibinfo
  {pages} {27537} (\bibinfo {year} {2013})}\BibitemShut {NoStop}%
\bibitem [{\citenamefont {Djurišić}\ and\ \citenamefont
  {Li}(1999)}]{Djurisix1999JAP}%
  \BibitemOpen
  \bibfield  {author} {\bibinfo {author} {\bibfnamefont {A.~B.}\ \bibnamefont
  {Djurišić}}\ and\ \bibinfo {author} {\bibfnamefont {E.~H.}\ \bibnamefont
  {Li}},\ }\bibfield  {title} {\bibinfo {title} {Optical properties of
  graphite},\ }\href {https://doi.org/10.1063/1.369370} {\bibfield  {journal}
  {\bibinfo  {journal} {Journal of Applied Physics}\ }\textbf {\bibinfo
  {volume} {85}},\ \bibinfo {pages} {7404} (\bibinfo {year}
  {1999})}\BibitemShut {NoStop}%
\bibitem [{\citenamefont {Tikuišis}\ \emph {et~al.}(2023)\citenamefont
  {Tikuišis}, \citenamefont {Dubroka}, \citenamefont {Uhlířová},
  \citenamefont {Speck}, \citenamefont {Seyller}, \citenamefont {Losurdo},
  \citenamefont {Orlita},\ and\ \citenamefont {Veis}}]{Tikuisis2023PRM}%
  \BibitemOpen
  \bibfield  {author} {\bibinfo {author} {\bibfnamefont {K.~K.}\ \bibnamefont
  {Tikuišis}}, \bibinfo {author} {\bibfnamefont {A.}~\bibnamefont {Dubroka}},
  \bibinfo {author} {\bibfnamefont {K.}~\bibnamefont {Uhlířová}}, \bibinfo
  {author} {\bibfnamefont {F.}~\bibnamefont {Speck}}, \bibinfo {author}
  {\bibfnamefont {T.}~\bibnamefont {Seyller}}, \bibinfo {author} {\bibfnamefont
  {M.}~\bibnamefont {Losurdo}}, \bibinfo {author} {\bibfnamefont
  {M.}~\bibnamefont {Orlita}},\ and\ \bibinfo {author} {\bibfnamefont
  {M.}~\bibnamefont {Veis}},\ }\bibfield  {title} {\bibinfo {title} {Dielectric
  function of epitaxial quasi-freestanding monolayer graphene on si-face 6h-sic
  in a broad spectral range},\ }\bibfield  {journal} {\bibinfo  {journal}
  {Physical Review Materials}\ }\textbf {\bibinfo {volume} {7}},\ \href
  {https://doi.org/10.1103/PhysRevMaterials.7.044201}
  {10.1103/PhysRevMaterials.7.044201} (\bibinfo {year} {2023})\BibitemShut
  {NoStop}%
\bibitem [{\citenamefont {{Rodríguez-de Marcos}}\ \emph
  {et~al.}(2016)\citenamefont {{Rodríguez-de Marcos}}, \citenamefont
  {Larruquert}, \citenamefont {Méndez},\ and\ \citenamefont
  {Aznárez}}]{Rodriguez2016OME}%
  \BibitemOpen
  \bibfield  {author} {\bibinfo {author} {\bibfnamefont {L.~V.}\ \bibnamefont
  {{Rodríguez-de Marcos}}}, \bibinfo {author} {\bibfnamefont {J.~I.}\
  \bibnamefont {Larruquert}}, \bibinfo {author} {\bibfnamefont {J.~A.}\
  \bibnamefont {Méndez}},\ and\ \bibinfo {author} {\bibfnamefont {J.~A.}\
  \bibnamefont {Aznárez}},\ }\bibfield  {title} {\bibinfo {title}
  {Self-consistent optical constants of sio$_2$ and ta$_2$o$_5$ films},\ }\href
  {https://doi.org/10.1364/ome.6.003622} {\bibfield  {journal} {\bibinfo
  {journal} {Optical Materials Express}\ }\textbf {\bibinfo {volume} {6}},\
  \bibinfo {pages} {3622} (\bibinfo {year} {2016})}\BibitemShut {NoStop}%
\bibitem [{\citenamefont {Sarkar}\ \emph {et~al.}(2019)\citenamefont {Sarkar},
  \citenamefont {Gupta}, \citenamefont {Kumar}, \citenamefont {Schubert},
  \citenamefont {Probst}, \citenamefont {Joseph},\ and\ \citenamefont
  {König}}]{Sarkar2019ACSAMI}%
  \BibitemOpen
  \bibfield  {author} {\bibinfo {author} {\bibfnamefont {S.}~\bibnamefont
  {Sarkar}}, \bibinfo {author} {\bibfnamefont {V.}~\bibnamefont {Gupta}},
  \bibinfo {author} {\bibfnamefont {M.}~\bibnamefont {Kumar}}, \bibinfo
  {author} {\bibfnamefont {J.}~\bibnamefont {Schubert}}, \bibinfo {author}
  {\bibfnamefont {P.~T.}\ \bibnamefont {Probst}}, \bibinfo {author}
  {\bibfnamefont {J.}~\bibnamefont {Joseph}},\ and\ \bibinfo {author}
  {\bibfnamefont {T.~A.}\ \bibnamefont {König}},\ }\bibfield  {title}
  {\bibinfo {title} {Hybridized guided-mode resonances via colloidal plasmonic
  self-assembled grating},\ }\href {https://doi.org/10.1021/acsami.8b20535}
  {\bibfield  {journal} {\bibinfo  {journal} {ACS Applied Materials and
  Interfaces}\ }\textbf {\bibinfo {volume} {11}},\ \bibinfo {pages} {13752}
  (\bibinfo {year} {2019})}\BibitemShut {NoStop}%
\bibitem [{\citenamefont {Bright}\ \emph {et~al.}(2013)\citenamefont {Bright},
  \citenamefont {Watjen}, \citenamefont {Zhang}, \citenamefont {Muratore},
  \citenamefont {Voevodin}, \citenamefont {Koukis}, \citenamefont {Tanner},\
  and\ \citenamefont {Arenas}}]{Bright2013JAP}%
  \BibitemOpen
  \bibfield  {author} {\bibinfo {author} {\bibfnamefont {T.~J.}\ \bibnamefont
  {Bright}}, \bibinfo {author} {\bibfnamefont {J.~I.}\ \bibnamefont {Watjen}},
  \bibinfo {author} {\bibfnamefont {Z.~M.}\ \bibnamefont {Zhang}}, \bibinfo
  {author} {\bibfnamefont {C.}~\bibnamefont {Muratore}}, \bibinfo {author}
  {\bibfnamefont {A.~A.}\ \bibnamefont {Voevodin}}, \bibinfo {author}
  {\bibfnamefont {D.~I.}\ \bibnamefont {Koukis}}, \bibinfo {author}
  {\bibfnamefont {D.~B.}\ \bibnamefont {Tanner}},\ and\ \bibinfo {author}
  {\bibfnamefont {D.~J.}\ \bibnamefont {Arenas}},\ }\bibfield  {title}
  {\bibinfo {title} {Infrared optical properties of amorphous and
  nanocrystalline ta$_2$o$_5$ thin films},\ }\bibfield  {journal} {\bibinfo
  {journal} {Journal of Applied Physics}\ }\textbf {\bibinfo {volume} {114}},\
  \href {https://doi.org/10.1063/1.4819325} {10.1063/1.4819325} (\bibinfo
  {year} {2013})\BibitemShut {NoStop}%
\bibitem [{\citenamefont {Aguilar}\ \emph {et~al.}(2019)\citenamefont
  {Aguilar}, \citenamefont {de~Castro}, \citenamefont {Godoy},\ and\
  \citenamefont {Dias}}]{Aguilar2019OME}%
  \BibitemOpen
  \bibfield  {author} {\bibinfo {author} {\bibfnamefont {O.}~\bibnamefont
  {Aguilar}}, \bibinfo {author} {\bibfnamefont {S.}~\bibnamefont {de~Castro}},
  \bibinfo {author} {\bibfnamefont {M.~P.~F.}\ \bibnamefont {Godoy}},\ and\
  \bibinfo {author} {\bibfnamefont {M.~R.~S.}\ \bibnamefont {Dias}},\
  }\bibfield  {title} {\bibinfo {title} {Optoelectronic characterization of
  zn$_{1-x}$cd$_x$o thin films as an alternative to photonic crystals in
  organic solar cells},\ }\href {https://doi.org/10.1364/ome.9.003638}
  {\bibfield  {journal} {\bibinfo  {journal} {Optical Materials Express}\
  }\textbf {\bibinfo {volume} {9}},\ \bibinfo {pages} {3638} (\bibinfo {year}
  {2019})}\BibitemShut {NoStop}%
\bibitem [{\citenamefont {Zelmon}\ \emph {et~al.}(1997)\citenamefont {Zelmon},
  \citenamefont {Small},\ and\ \citenamefont {Jundt}}]{Zelman1997JOSAB}%
  \BibitemOpen
  \bibfield  {author} {\bibinfo {author} {\bibfnamefont {D.~E.}\ \bibnamefont
  {Zelmon}}, \bibinfo {author} {\bibfnamefont {D.~L.}\ \bibnamefont {Small}},\
  and\ \bibinfo {author} {\bibfnamefont {D.}~\bibnamefont {Jundt}},\ }\bibfield
   {title} {\bibinfo {title} {Infrared corrected sellmeier coefficients for
  congruently grown lithium niobate and 5 mol. \% magnesium oxide-doped lithium
  niobate},\ }\href@noop {} {\bibfield  {journal} {\bibinfo  {journal} {Journal
  of the Optical Society of America B}\ }\textbf {\bibinfo {volume} {14}},\
  \bibinfo {pages} {3319} (\bibinfo {year} {1997})}\BibitemShut {NoStop}%
\bibitem [{\citenamefont {Stefaniuk}\ \emph {et~al.}(2023)\citenamefont
  {Stefaniuk}, \citenamefont {Suffczyński}, \citenamefont {Wierzbowska},
  \citenamefont {Domagała}, \citenamefont {Kisielewski}, \citenamefont
  {Kłos}, \citenamefont {Korneluk},\ and\ \citenamefont
  {Teisseyre}}]{Stefaniuk2023PRB}%
  \BibitemOpen
  \bibfield  {author} {\bibinfo {author} {\bibfnamefont {T.}~\bibnamefont
  {Stefaniuk}}, \bibinfo {author} {\bibfnamefont {J.}~\bibnamefont
  {Suffczyński}}, \bibinfo {author} {\bibfnamefont {M.}~\bibnamefont
  {Wierzbowska}}, \bibinfo {author} {\bibfnamefont {J.~Z.}\ \bibnamefont
  {Domagała}}, \bibinfo {author} {\bibfnamefont {J.}~\bibnamefont
  {Kisielewski}}, \bibinfo {author} {\bibfnamefont {A.}~\bibnamefont {Kłos}},
  \bibinfo {author} {\bibfnamefont {A.}~\bibnamefont {Korneluk}},\ and\
  \bibinfo {author} {\bibfnamefont {H.}~\bibnamefont {Teisseyre}},\ }\bibfield
  {title} {\bibinfo {title} {Optical, electronic, and structural properties of
  {ScAlMgO$_4$}},\ }\bibfield  {journal} {\bibinfo  {journal} {Physical Review
  B}\ }\textbf {\bibinfo {volume} {107}},\ \href
  {https://doi.org/10.1103/PhysRevB.107.085205} {10.1103/PhysRevB.107.085205}
  (\bibinfo {year} {2023})\BibitemShut {NoStop}%
\bibitem [{\citenamefont {Beliaev}\ \emph {et~al.}(2022)\citenamefont
  {Beliaev}, \citenamefont {Shkondin}, \citenamefont {Lavrinenko},\ and\
  \citenamefont {Takayama}}]{Beliaev2022TSF}%
  \BibitemOpen
  \bibfield  {author} {\bibinfo {author} {\bibfnamefont {L.~Y.}\ \bibnamefont
  {Beliaev}}, \bibinfo {author} {\bibfnamefont {E.}~\bibnamefont {Shkondin}},
  \bibinfo {author} {\bibfnamefont {A.~V.}\ \bibnamefont {Lavrinenko}},\ and\
  \bibinfo {author} {\bibfnamefont {O.}~\bibnamefont {Takayama}},\ }\bibfield
  {title} {\bibinfo {title} {Optical, structural and composition properties of
  silicon nitride films deposited by reactive radio-frequency sputtering, low
  pressure and plasma-enhanced chemical vapor deposition},\ }\bibfield
  {journal} {\bibinfo  {journal} {Thin Solid Films}\ }\textbf {\bibinfo
  {volume} {763}},\ \href {https://doi.org/10.1016/j.tsf.2022.139568}
  {10.1016/j.tsf.2022.139568} (\bibinfo {year} {2022})\BibitemShut {NoStop}%
\bibitem [{\citenamefont {Grudinin}\ \emph {et~al.}(2023)\citenamefont
  {Grudinin}, \citenamefont {Ermolaev}, \citenamefont {Baranov}, \citenamefont
  {Toksumakov}, \citenamefont {Voronin}, \citenamefont {Slavich}, \citenamefont
  {Vyshnevyy}, \citenamefont {Mazitov}, \citenamefont {Kruglov}, \citenamefont
  {Ghazaryan}, \citenamefont {Arsenin}, \citenamefont {Novoselov},\ and\
  \citenamefont {Volkov}}]{Grudinin2023MH}%
  \BibitemOpen
  \bibfield  {author} {\bibinfo {author} {\bibfnamefont {D.~V.}\ \bibnamefont
  {Grudinin}}, \bibinfo {author} {\bibfnamefont {G.~A.}\ \bibnamefont
  {Ermolaev}}, \bibinfo {author} {\bibfnamefont {D.~G.}\ \bibnamefont
  {Baranov}}, \bibinfo {author} {\bibfnamefont {A.~N.}\ \bibnamefont
  {Toksumakov}}, \bibinfo {author} {\bibfnamefont {K.~V.}\ \bibnamefont
  {Voronin}}, \bibinfo {author} {\bibfnamefont {A.~S.}\ \bibnamefont
  {Slavich}}, \bibinfo {author} {\bibfnamefont {A.~A.}\ \bibnamefont
  {Vyshnevyy}}, \bibinfo {author} {\bibfnamefont {A.~B.}\ \bibnamefont
  {Mazitov}}, \bibinfo {author} {\bibfnamefont {I.~A.}\ \bibnamefont
  {Kruglov}}, \bibinfo {author} {\bibfnamefont {D.~A.}\ \bibnamefont
  {Ghazaryan}}, \bibinfo {author} {\bibfnamefont {A.~V.}\ \bibnamefont
  {Arsenin}}, \bibinfo {author} {\bibfnamefont {K.~S.}\ \bibnamefont
  {Novoselov}},\ and\ \bibinfo {author} {\bibfnamefont {V.~S.}\ \bibnamefont
  {Volkov}},\ }\bibfield  {title} {\bibinfo {title} {Hexagonal boron nitride
  nanophotonics: a record-breaking material for the ultraviolet and visible
  spectral ranges},\ }\href {https://doi.org/10.1039/d3mh00215b} {\bibfield
  {journal} {\bibinfo  {journal} {Materials Horizons}\ }\textbf {\bibinfo
  {volume} {10}},\ \bibinfo {pages} {2427} (\bibinfo {year}
  {2023})}\BibitemShut {NoStop}%
\bibitem [{\citenamefont {Joseph}\ \emph {et~al.}(2020)\citenamefont {Joseph},
  \citenamefont {Sarkar},\ and\ \citenamefont {Joseph}}]{Joseph2020ACSAMI}%
  \BibitemOpen
  \bibfield  {author} {\bibinfo {author} {\bibfnamefont {S.}~\bibnamefont
  {Joseph}}, \bibinfo {author} {\bibfnamefont {S.}~\bibnamefont {Sarkar}},\
  and\ \bibinfo {author} {\bibfnamefont {J.}~\bibnamefont {Joseph}},\
  }\bibfield  {title} {\bibinfo {title} {Grating-coupled surface
  plasmon-polariton sensing at a flat metal--analyte interface in a
  hybrid-configuration},\ }\href@noop {} {\bibfield  {journal} {\bibinfo
  {journal} {ACS Applied Materials \& Interfaces}\ }\textbf {\bibinfo {volume}
  {12}},\ \bibinfo {pages} {46519} (\bibinfo {year} {2020})}\BibitemShut
  {NoStop}%
\bibitem [{\citenamefont {Fang}\ \emph {et~al.}(2020)\citenamefont {Fang},
  \citenamefont {Wang}, \citenamefont {Gu}, \citenamefont {Tong}, \citenamefont
  {Song}, \citenamefont {Xie}, \citenamefont {Zhou}, \citenamefont {Chen},
  \citenamefont {Jiang}, \citenamefont {Jiang},\ and\ \citenamefont
  {Liu}}]{Fang2020ASS}%
  \BibitemOpen
  \bibfield  {author} {\bibinfo {author} {\bibfnamefont {M.}~\bibnamefont
  {Fang}}, \bibinfo {author} {\bibfnamefont {Z.}~\bibnamefont {Wang}}, \bibinfo
  {author} {\bibfnamefont {H.}~\bibnamefont {Gu}}, \bibinfo {author}
  {\bibfnamefont {M.}~\bibnamefont {Tong}}, \bibinfo {author} {\bibfnamefont
  {B.}~\bibnamefont {Song}}, \bibinfo {author} {\bibfnamefont {X.}~\bibnamefont
  {Xie}}, \bibinfo {author} {\bibfnamefont {T.}~\bibnamefont {Zhou}}, \bibinfo
  {author} {\bibfnamefont {X.}~\bibnamefont {Chen}}, \bibinfo {author}
  {\bibfnamefont {H.}~\bibnamefont {Jiang}}, \bibinfo {author} {\bibfnamefont
  {T.}~\bibnamefont {Jiang}},\ and\ \bibinfo {author} {\bibfnamefont
  {S.}~\bibnamefont {Liu}},\ }\bibfield  {title} {\bibinfo {title}
  {Layer-dependent dielectric permittivity of topological insulator
  bi$_2$se$_3$ thin films},\ }\bibfield  {journal} {\bibinfo  {journal}
  {Applied Surface Science}\ }\textbf {\bibinfo {volume} {509}},\ \href
  {https://doi.org/10.1016/j.apsusc.2019.144822} {10.1016/j.apsusc.2019.144822}
  (\bibinfo {year} {2020})\BibitemShut {NoStop}%
\bibitem [{\citenamefont {Querry}(1987)}]{Querry1987CR}%
  \BibitemOpen
  \bibfield  {author} {\bibinfo {author} {\bibfnamefont {M.~R.}\ \bibnamefont
  {Querry}},\ }\href@noop {} {\bibinfo {title} {Optical constants of minerals
  and other materials from the millimeter to the ultraviolet,}} (\bibinfo
  {year} {1987})\BibitemShut {NoStop}%
\bibitem [{\citenamefont {Treharne}\ \emph {et~al.}(2011)\citenamefont
  {Treharne}, \citenamefont {Seymour-Pierce}, \citenamefont {Durose},
  \citenamefont {Hutchings}, \citenamefont {Roncallo},\ and\ \citenamefont
  {Lane}}]{Treharne2011JPCS}%
  \BibitemOpen
  \bibfield  {author} {\bibinfo {author} {\bibfnamefont {R.~E.}\ \bibnamefont
  {Treharne}}, \bibinfo {author} {\bibfnamefont {A.}~\bibnamefont
  {Seymour-Pierce}}, \bibinfo {author} {\bibfnamefont {K.}~\bibnamefont
  {Durose}}, \bibinfo {author} {\bibfnamefont {K.}~\bibnamefont {Hutchings}},
  \bibinfo {author} {\bibfnamefont {S.}~\bibnamefont {Roncallo}},\ and\
  \bibinfo {author} {\bibfnamefont {D.}~\bibnamefont {Lane}},\ }\bibfield
  {title} {\bibinfo {title} {Optical design and fabrication of fully sputtered
  cdte/cds solar cells},\ }\bibfield  {journal} {\bibinfo  {journal} {Journal
  of Physics: Conference Series}\ }\textbf {\bibinfo {volume} {286}},\ \href
  {https://doi.org/10.1088/1742-6596/286/1/012038}
  {10.1088/1742-6596/286/1/012038} (\bibinfo {year} {2011})\BibitemShut
  {NoStop}%
\bibitem [{\citenamefont {Gissibl}\ \emph {et~al.}(2017)\citenamefont
  {Gissibl}, \citenamefont {Wagner}, \citenamefont {Sykora}, \citenamefont
  {Schmid},\ and\ \citenamefont {Giessen}}]{Gissibl2017OME}%
  \BibitemOpen
  \bibfield  {author} {\bibinfo {author} {\bibfnamefont {T.}~\bibnamefont
  {Gissibl}}, \bibinfo {author} {\bibfnamefont {S.}~\bibnamefont {Wagner}},
  \bibinfo {author} {\bibfnamefont {J.}~\bibnamefont {Sykora}}, \bibinfo
  {author} {\bibfnamefont {M.}~\bibnamefont {Schmid}},\ and\ \bibinfo {author}
  {\bibfnamefont {H.}~\bibnamefont {Giessen}},\ }\bibfield  {title} {\bibinfo
  {title} {Refractive index measurements of photo-resists for three-dimensional
  direct laser writing},\ }\href@noop {} {\bibfield  {journal} {\bibinfo
  {journal} {Optical Materials Express}\ }\textbf {\bibinfo {volume} {7}},\
  \bibinfo {pages} {2293} (\bibinfo {year} {2017})}\BibitemShut {NoStop}%
\bibitem [{ZEO(2010)}]{ZEONreport}%
  \BibitemOpen
  \href@noop {} {\emph {\bibinfo {title} {ZEP520A Technical Report}}},\
  \bibinfo {type} {Tech. Rep.}\ (\bibinfo  {institution} {ZEON CORPORATION
  Electronics Materials Division},\ \bibinfo {year} {2010})\BibitemShut
  {NoStop}%
\bibitem [{\citenamefont {Munkhbat}\ \emph {et~al.}(2022)\citenamefont
  {Munkhbat}, \citenamefont {Wróbel}, \citenamefont {Antosiewicz},\ and\
  \citenamefont {Shegai}}]{Munkhbat2022ACSPh}%
  \BibitemOpen
  \bibfield  {author} {\bibinfo {author} {\bibfnamefont {B.}~\bibnamefont
  {Munkhbat}}, \bibinfo {author} {\bibfnamefont {P.}~\bibnamefont {Wróbel}},
  \bibinfo {author} {\bibfnamefont {T.~J.}\ \bibnamefont {Antosiewicz}},\ and\
  \bibinfo {author} {\bibfnamefont {T.~O.}\ \bibnamefont {Shegai}},\ }\bibfield
   {title} {\bibinfo {title} {Optical constants of several multilayer
  transition metal dichalcogenides measured by spectroscopic ellipsometry in
  the 300–1700 nm range: High index, anisotropy, and hyperbolicity},\ }\href
  {https://doi.org/10.1021/acsphotonics.2c00433} {\bibfield  {journal}
  {\bibinfo  {journal} {ACS Photonics}\ }\textbf {\bibinfo {volume} {9}},\
  \bibinfo {pages} {2398} (\bibinfo {year} {2022})}\BibitemShut {NoStop}%
\bibitem [{\citenamefont {Ermolaev}\ \emph {et~al.}(2021)\citenamefont
  {Ermolaev}, \citenamefont {Voronin}, \citenamefont {Tatmyshevskiy},
  \citenamefont {Mazitov}, \citenamefont {Slavich}, \citenamefont {Yakubovsky},
  \citenamefont {Tselin}, \citenamefont {Mironov}, \citenamefont {Romanov},
  \citenamefont {Markeev}, \citenamefont {Kruglov}, \citenamefont {Novikov},
  \citenamefont {Vyshnevyy}, \citenamefont {Arsenin},\ and\ \citenamefont
  {Volkov}}]{Ermolaev2021Nm}%
  \BibitemOpen
  \bibfield  {author} {\bibinfo {author} {\bibfnamefont {G.~A.}\ \bibnamefont
  {Ermolaev}}, \bibinfo {author} {\bibfnamefont {K.~V.}\ \bibnamefont
  {Voronin}}, \bibinfo {author} {\bibfnamefont {M.~K.}\ \bibnamefont
  {Tatmyshevskiy}}, \bibinfo {author} {\bibfnamefont {A.~B.}\ \bibnamefont
  {Mazitov}}, \bibinfo {author} {\bibfnamefont {A.~S.}\ \bibnamefont
  {Slavich}}, \bibinfo {author} {\bibfnamefont {D.~I.}\ \bibnamefont
  {Yakubovsky}}, \bibinfo {author} {\bibfnamefont {A.~P.}\ \bibnamefont
  {Tselin}}, \bibinfo {author} {\bibfnamefont {M.~S.}\ \bibnamefont {Mironov}},
  \bibinfo {author} {\bibfnamefont {R.~I.}\ \bibnamefont {Romanov}}, \bibinfo
  {author} {\bibfnamefont {A.~M.}\ \bibnamefont {Markeev}}, \bibinfo {author}
  {\bibfnamefont {I.~A.}\ \bibnamefont {Kruglov}}, \bibinfo {author}
  {\bibfnamefont {S.~M.}\ \bibnamefont {Novikov}}, \bibinfo {author}
  {\bibfnamefont {A.~A.}\ \bibnamefont {Vyshnevyy}}, \bibinfo {author}
  {\bibfnamefont {A.~V.}\ \bibnamefont {Arsenin}},\ and\ \bibinfo {author}
  {\bibfnamefont {V.~S.}\ \bibnamefont {Volkov}},\ }\bibfield  {title}
  {\bibinfo {title} {Broadband optical properties of atomically thin pts$_2$
  and ptse$_2$},\ }\bibfield  {journal} {\bibinfo  {journal} {Nanomaterials}\
  }\textbf {\bibinfo {volume} {11}},\ \href
  {https://doi.org/10.3390/nano11123269} {10.3390/nano11123269} (\bibinfo
  {year} {2021})\BibitemShut {NoStop}%
\bibitem [{\citenamefont {Polyanskiy}(2024)}]{Polyanskiy2024SD}%
  \BibitemOpen
  \bibfield  {author} {\bibinfo {author} {\bibfnamefont {M.~N.}\ \bibnamefont
  {Polyanskiy}},\ }\bibfield  {title} {\bibinfo {title} {Refractiveindex.info
  database of optical constants},\ }\bibfield  {journal} {\bibinfo  {journal}
  {Scientific Data}\ }\textbf {\bibinfo {volume} {11}},\ \href
  {https://doi.org/10.1038/s41597-023-02898-2} {10.1038/s41597-023-02898-2}
  (\bibinfo {year} {2024})\BibitemShut {NoStop}%
\bibitem [{\citenamefont {Mateus}\ \emph {et~al.}(2004)\citenamefont {Mateus},
  \citenamefont {Huang}, \citenamefont {Chen}, \citenamefont {Chang-Hasnain},\
  and\ \citenamefont {Suzuki}}]{Mateus2007IEEEPTL}%
  \BibitemOpen
  \bibfield  {author} {\bibinfo {author} {\bibfnamefont {C.~F.~R.}\
  \bibnamefont {Mateus}}, \bibinfo {author} {\bibfnamefont {M.~C.~Y.}\
  \bibnamefont {Huang}}, \bibinfo {author} {\bibfnamefont {L.}~\bibnamefont
  {Chen}}, \bibinfo {author} {\bibfnamefont {C.~J.}\ \bibnamefont
  {Chang-Hasnain}},\ and\ \bibinfo {author} {\bibfnamefont {Y.}~\bibnamefont
  {Suzuki}},\ }\bibfield  {title} {\bibinfo {title} {Broad-band mirror
  (1.12-1.62 \textmu m) using a subwavelength grating},\ }\href@noop {}
  {\bibfield  {journal} {\bibinfo  {journal} {IEEE Photonics Technology
  Letters}\ }\textbf {\bibinfo {volume} {16}},\ \bibinfo {pages} {1676}
  (\bibinfo {year} {2004})}\BibitemShut {NoStop}%
\bibitem [{\citenamefont {Zhou}\ \emph {et~al.}(2008)\citenamefont {Zhou},
  \citenamefont {Moewe}, \citenamefont {Kern}, \citenamefont {Huang},\ and\
  \citenamefont {Chang-Hasnain}}]{Zhou2008OE}%
  \BibitemOpen
  \bibfield  {author} {\bibinfo {author} {\bibfnamefont {Y.}~\bibnamefont
  {Zhou}}, \bibinfo {author} {\bibfnamefont {M.}~\bibnamefont {Moewe}},
  \bibinfo {author} {\bibfnamefont {J.}~\bibnamefont {Kern}}, \bibinfo {author}
  {\bibfnamefont {M.~C.}\ \bibnamefont {Huang}},\ and\ \bibinfo {author}
  {\bibfnamefont {C.~J.}\ \bibnamefont {Chang-Hasnain}},\ }\bibfield  {title}
  {\bibinfo {title} {Ultrahigh-q nanocavities in two-dimensional photonic
  crystal slabs},\ }\href@noop {} {\bibfield  {journal} {\bibinfo  {journal}
  {Optics Express}\ }\textbf {\bibinfo {volume} {16}},\ \bibinfo {pages}
  {17282} (\bibinfo {year} {2008})}\BibitemShut {NoStop}%
\bibitem [{\citenamefont {Chase}\ \emph {et~al.}(2010)\citenamefont {Chase},
  \citenamefont {Rao}, \citenamefont {Hofmann},\ and\ \citenamefont
  {Chang-Hasnain}}]{Chase2010OE}%
  \BibitemOpen
  \bibfield  {author} {\bibinfo {author} {\bibfnamefont {C.}~\bibnamefont
  {Chase}}, \bibinfo {author} {\bibfnamefont {Y.}~\bibnamefont {Rao}}, \bibinfo
  {author} {\bibfnamefont {W.}~\bibnamefont {Hofmann}},\ and\ \bibinfo {author}
  {\bibfnamefont {C.~J.}\ \bibnamefont {Chang-Hasnain}},\ }\bibfield  {title}
  {\bibinfo {title} {550 nm high contrast grating vcsel},\ }\href@noop {}
  {\bibfield  {journal} {\bibinfo  {journal} {Optics Express}\ }\textbf
  {\bibinfo {volume} {18}},\ \bibinfo {pages} {15461} (\bibinfo {year}
  {2010})}\BibitemShut {NoStop}%
\bibitem [{\citenamefont {Ansbaek}\ \emph {et~al.}(2013)\citenamefont
  {Ansbaek}, \citenamefont {Chung}, \citenamefont {Semenova},\ and\
  \citenamefont {Yvind}}]{Ansbaek2013IEEEPTL}%
  \BibitemOpen
  \bibfield  {author} {\bibinfo {author} {\bibfnamefont {T.}~\bibnamefont
  {Ansbaek}}, \bibinfo {author} {\bibfnamefont {I.~S.}\ \bibnamefont {Chung}},
  \bibinfo {author} {\bibfnamefont {E.~S.}\ \bibnamefont {Semenova}},\ and\
  \bibinfo {author} {\bibfnamefont {K.}~\bibnamefont {Yvind}},\ }\bibfield
  {title} {\bibinfo {title} {1060-nm tunable monolithic high index contrast
  subwavelength grating vcsel},\ }\href
  {https://doi.org/10.1109/LPT.2012.2236087} {\bibfield  {journal} {\bibinfo
  {journal} {IEEE Photonics Technology Letters}\ }\textbf {\bibinfo {volume}
  {25}},\ \bibinfo {pages} {365} (\bibinfo {year} {2013})}\BibitemShut
  {NoStop}%
\bibitem [{\citenamefont {Zhang}\ \emph {et~al.}(2020)\citenamefont {Zhang},
  \citenamefont {Zhang}, \citenamefont {Gogna}, \citenamefont {Chen},\ and\
  \citenamefont {Deng}}]{Zhang2020SSC}%
  \BibitemOpen
  \bibfield  {author} {\bibinfo {author} {\bibfnamefont {Z.}~\bibnamefont
  {Zhang}}, \bibinfo {author} {\bibfnamefont {L.}~\bibnamefont {Zhang}},
  \bibinfo {author} {\bibfnamefont {R.}~\bibnamefont {Gogna}}, \bibinfo
  {author} {\bibfnamefont {Z.}~\bibnamefont {Chen}},\ and\ \bibinfo {author}
  {\bibfnamefont {H.}~\bibnamefont {Deng}},\ }\bibfield  {title} {\bibinfo
  {title} {Large enhancement of second-harmonic generation in mos$_2$ by one
  dimensional photonic crystals},\ }\bibfield  {journal} {\bibinfo  {journal}
  {Solid State Communications}\ }\textbf {\bibinfo {volume} {322}},\ \href
  {https://doi.org/10.1016/j.ssc.2020.114043} {10.1016/j.ssc.2020.114043}
  (\bibinfo {year} {2020})\BibitemShut {NoStop}%
\bibitem [{\citenamefont {Marciniak}\ \emph {et~al.}(2021)\citenamefont
  {Marciniak}, \citenamefont {Chang}, \citenamefont {Lu}, \citenamefont
  {Hjort}, \citenamefont {Åsa Haglund}, \citenamefont {Łucja Marona},
  \citenamefont {Gramala}, \citenamefont {Modrzyński}, \citenamefont
  {Kudrawiec}, \citenamefont {Sawicki}, \citenamefont {Bożek}, \citenamefont
  {Pacuski}, \citenamefont {Suffczyński}, \citenamefont {G\k{}bski},
  \citenamefont {Broda}, \citenamefont {Muszalski}, \citenamefont {Lott},\ and\
  \citenamefont {Czyszanowski}}]{Marciniak2021ACSP}%
  \BibitemOpen
  \bibfield  {author} {\bibinfo {author} {\bibfnamefont {M.}~\bibnamefont
  {Marciniak}}, \bibinfo {author} {\bibfnamefont {T.~S.}\ \bibnamefont
  {Chang}}, \bibinfo {author} {\bibfnamefont {T.~C.}\ \bibnamefont {Lu}},
  \bibinfo {author} {\bibfnamefont {F.}~\bibnamefont {Hjort}}, \bibinfo
  {author} {\bibnamefont {Åsa Haglund}}, \bibinfo {author} {\bibnamefont
  {Łucja Marona}}, \bibinfo {author} {\bibfnamefont {M.}~\bibnamefont
  {Gramala}}, \bibinfo {author} {\bibfnamefont {P.}~\bibnamefont
  {Modrzyński}}, \bibinfo {author} {\bibfnamefont {R.}~\bibnamefont
  {Kudrawiec}}, \bibinfo {author} {\bibfnamefont {K.}~\bibnamefont {Sawicki}},
  \bibinfo {author} {\bibfnamefont {R.}~\bibnamefont {Bożek}}, \bibinfo
  {author} {\bibfnamefont {W.}~\bibnamefont {Pacuski}}, \bibinfo {author}
  {\bibfnamefont {J.}~\bibnamefont {Suffczyński}}, \bibinfo {author}
  {\bibfnamefont {M.}~\bibnamefont {G\k{}bski}}, \bibinfo {author}
  {\bibfnamefont {A.}~\bibnamefont {Broda}}, \bibinfo {author} {\bibfnamefont
  {J.}~\bibnamefont {Muszalski}}, \bibinfo {author} {\bibfnamefont {J.~A.}\
  \bibnamefont {Lott}},\ and\ \bibinfo {author} {\bibfnamefont
  {T.}~\bibnamefont {Czyszanowski}},\ }\bibfield  {title} {\bibinfo {title}
  {Impact of stripe shape on the reflectivity of monolithic high contrast
  gratings},\ }\href {https://doi.org/10.1021/acsphotonics.1c00850} {\bibfield
  {journal} {\bibinfo  {journal} {ACS Photonics}\ }\textbf {\bibinfo {volume}
  {8}},\ \bibinfo {pages} {3173} (\bibinfo {year} {2021})}\BibitemShut
  {NoStop}%
\bibitem [{\citenamefont {Lai}\ \emph {et~al.}(2015)\citenamefont {Lai},
  \citenamefont {Matsutani}, \citenamefont {Lu}, \citenamefont {Wang},\ and\
  \citenamefont {Koyama}}]{Lai2015procSPIE}%
  \BibitemOpen
  \bibfield  {author} {\bibinfo {author} {\bibfnamefont {Y.-Y.}\ \bibnamefont
  {Lai}}, \bibinfo {author} {\bibfnamefont {A.}~\bibnamefont {Matsutani}},
  \bibinfo {author} {\bibfnamefont {T.-C.}\ \bibnamefont {Lu}}, \bibinfo
  {author} {\bibfnamefont {S.-C.}\ \bibnamefont {Wang}},\ and\ \bibinfo
  {author} {\bibfnamefont {F.}~\bibnamefont {Koyama}},\ }\bibfield  {title}
  {\bibinfo {title} {Fabrication of sic membrane hcg blue reflector using
  nanoimprint lithography},\ }in\ \href {https://doi.org/10.1117/12.2077391}
  {\emph {\bibinfo {booktitle} {High Contrast Metastructures IV}}},\ Vol.\
  \bibinfo {volume} {9372}\ (\bibinfo  {publisher} {SPIE},\ \bibinfo {year}
  {2015})\ p.\ \bibinfo {pages} {937207}\BibitemShut {NoStop}%
\bibitem [{\citenamefont {Hogan}\ \emph {et~al.}(2016)\citenamefont {Hogan},
  \citenamefont {Hegarty}, \citenamefont {Lewis}, \citenamefont {Romero-Vivas},
  \citenamefont {Ochalski},\ and\ \citenamefont {Huyet}}]{Hogan2016OL}%
  \BibitemOpen
  \bibfield  {author} {\bibinfo {author} {\bibfnamefont {B.}~\bibnamefont
  {Hogan}}, \bibinfo {author} {\bibfnamefont {S.~P.}\ \bibnamefont {Hegarty}},
  \bibinfo {author} {\bibfnamefont {L.}~\bibnamefont {Lewis}}, \bibinfo
  {author} {\bibfnamefont {J.}~\bibnamefont {Romero-Vivas}}, \bibinfo {author}
  {\bibfnamefont {T.~J.}\ \bibnamefont {Ochalski}},\ and\ \bibinfo {author}
  {\bibfnamefont {G.}~\bibnamefont {Huyet}},\ }\bibfield  {title} {\bibinfo
  {title} {Realization of high-contrast gratings operating at 10 \textmu m},\
  }\href {https://doi.org/10.1364/ol.41.005130} {\bibfield  {journal} {\bibinfo
   {journal} {Optics Letters}\ }\textbf {\bibinfo {volume} {41}},\ \bibinfo
  {pages} {5130} (\bibinfo {year} {2016})}\BibitemShut {NoStop}%
\bibitem [{\citenamefont {Hashemi}\ \emph {et~al.}(2015)\citenamefont
  {Hashemi}, \citenamefont {Bengtsson}, \citenamefont {Gustavsson},
  \citenamefont {Carlsson}, \citenamefont {Rossbach},\ and\ \citenamefont {Åsa
  Haglund}}]{Hashemi2015JVST}%
  \BibitemOpen
  \bibfield  {author} {\bibinfo {author} {\bibfnamefont {E.}~\bibnamefont
  {Hashemi}}, \bibinfo {author} {\bibfnamefont {J.}~\bibnamefont {Bengtsson}},
  \bibinfo {author} {\bibfnamefont {J.~S.}\ \bibnamefont {Gustavsson}},
  \bibinfo {author} {\bibfnamefont {S.}~\bibnamefont {Carlsson}}, \bibinfo
  {author} {\bibfnamefont {G.}~\bibnamefont {Rossbach}},\ and\ \bibinfo
  {author} {\bibnamefont {Åsa Haglund}},\ }\bibfield  {title} {\bibinfo
  {title} {Tio$_2$ membrane high-contrast grating reflectors for
  vertical-cavity light-emitters in the visible wavelength regime},\ }\href
  {https://doi.org/10.1116/1.4929416} {\bibfield  {journal} {\bibinfo
  {journal} {Journal of Vacuum Science \& Technology B, Nanotechnology and
  Microelectronics: Materials, Processing, Measurement, and Phenomena}\
  }\textbf {\bibinfo {volume} {33}},\ \bibinfo {pages} {050603} (\bibinfo
  {year} {2015})}\BibitemShut {NoStop}%
\bibitem [{\citenamefont {Benimetskiy}\ \emph {et~al.}(2020)\citenamefont
  {Benimetskiy}, \citenamefont {Kravtsov}, \citenamefont {Khestanova},
  \citenamefont {Mukhin}, \citenamefont {Sinev}, \citenamefont {Samusev},
  \citenamefont {Shelykh}, \citenamefont {Krizhanovskii}, \citenamefont
  {Skolnick},\ and\ \citenamefont {Iorsh}}]{Benimetskiy2020JPCS}%
  \BibitemOpen
  \bibfield  {author} {\bibinfo {author} {\bibfnamefont {F.~A.}\ \bibnamefont
  {Benimetskiy}}, \bibinfo {author} {\bibfnamefont {V.}~\bibnamefont
  {Kravtsov}}, \bibinfo {author} {\bibfnamefont {E.}~\bibnamefont
  {Khestanova}}, \bibinfo {author} {\bibfnamefont {I.~S.}\ \bibnamefont
  {Mukhin}}, \bibinfo {author} {\bibfnamefont {I.~S.}\ \bibnamefont {Sinev}},
  \bibinfo {author} {\bibfnamefont {A.~K.}\ \bibnamefont {Samusev}}, \bibinfo
  {author} {\bibfnamefont {I.~A.}\ \bibnamefont {Shelykh}}, \bibinfo {author}
  {\bibfnamefont {D.~N.}\ \bibnamefont {Krizhanovskii}}, \bibinfo {author}
  {\bibfnamefont {M.~S.}\ \bibnamefont {Skolnick}},\ and\ \bibinfo {author}
  {\bibfnamefont {I.~V.}\ \bibnamefont {Iorsh}},\ }\bibfield  {title} {\bibinfo
  {title} {Strong coupling of excitons in 2d mose2/hbn heterostructure with
  optical bound states in the continuum},\ }\bibfield  {journal} {\bibinfo
  {journal} {Journal of Physics: Conference Series}\ }\textbf {\bibinfo
  {volume} {1461}},\ \href {https://doi.org/10.1088/1742-6596/1461/1/012012}
  {10.1088/1742-6596/1461/1/012012} (\bibinfo {year} {2020})\BibitemShut
  {NoStop}%
\bibitem [{\citenamefont {Joseph}\ \emph {et~al.}(2021)\citenamefont {Joseph},
  \citenamefont {Sarkar}, \citenamefont {Khan},\ and\ \citenamefont
  {Joseph}}]{Joseph2021AOM}%
  \BibitemOpen
  \bibfield  {author} {\bibinfo {author} {\bibfnamefont {S.}~\bibnamefont
  {Joseph}}, \bibinfo {author} {\bibfnamefont {S.}~\bibnamefont {Sarkar}},
  \bibinfo {author} {\bibfnamefont {S.}~\bibnamefont {Khan}},\ and\ \bibinfo
  {author} {\bibfnamefont {J.}~\bibnamefont {Joseph}},\ }\bibfield  {title}
  {\bibinfo {title} {Exploring the optical bound state in the continuum in a
  dielectric grating coupled plasmonic hybrid system},\ }\href@noop {}
  {\bibfield  {journal} {\bibinfo  {journal} {Advanced Optical Materials}\
  }\textbf {\bibinfo {volume} {9}},\ \bibinfo {pages} {2001895} (\bibinfo
  {year} {2021})}\BibitemShut {NoStop}%
\bibitem [{\citenamefont {Wang}\ \emph {et~al.}(2021)\citenamefont {Wang},
  \citenamefont {Fan}, \citenamefont {Zhang}, \citenamefont {Tang},
  \citenamefont {Song}, \citenamefont {Han},\ and\ \citenamefont
  {Xiao}}]{Wang2021ACSN}%
  \BibitemOpen
  \bibfield  {author} {\bibinfo {author} {\bibfnamefont {Y.}~\bibnamefont
  {Wang}}, \bibinfo {author} {\bibfnamefont {Y.}~\bibnamefont {Fan}}, \bibinfo
  {author} {\bibfnamefont {X.}~\bibnamefont {Zhang}}, \bibinfo {author}
  {\bibfnamefont {H.}~\bibnamefont {Tang}}, \bibinfo {author} {\bibfnamefont
  {Q.}~\bibnamefont {Song}}, \bibinfo {author} {\bibfnamefont {J.}~\bibnamefont
  {Han}},\ and\ \bibinfo {author} {\bibfnamefont {S.}~\bibnamefont {Xiao}},\
  }\bibfield  {title} {\bibinfo {title} {Highly controllable etchless
  perovskite microlasers based on bound states in the continuum},\ }\href@noop
  {} {\bibfield  {journal} {\bibinfo  {journal} {ACS Nano}\ }\textbf {\bibinfo
  {volume} {15}},\ \bibinfo {pages} {7386} (\bibinfo {year}
  {2021})}\BibitemShut {NoStop}%
\bibitem [{\citenamefont {Pruszyńska-Karbownik}\ \emph
  {et~al.}(2023)\citenamefont {Pruszyńska-Karbownik}, \citenamefont {Jandura},
  \citenamefont {Dems}, \citenamefont {Łukasz Zinkiewicz}, \citenamefont
  {Broda}, \citenamefont {Gębski}, \citenamefont {Muszalski}, \citenamefont
  {Pudiš}, \citenamefont {Suffczyński},\ and\ \citenamefont
  {Czyszanowski}}]{EPK2023NPh}%
  \BibitemOpen
  \bibfield  {author} {\bibinfo {author} {\bibfnamefont {E.}~\bibnamefont
  {Pruszyńska-Karbownik}}, \bibinfo {author} {\bibfnamefont {D.}~\bibnamefont
  {Jandura}}, \bibinfo {author} {\bibfnamefont {M.}~\bibnamefont {Dems}},
  \bibinfo {author} {\bibnamefont {Łukasz Zinkiewicz}}, \bibinfo {author}
  {\bibfnamefont {A.}~\bibnamefont {Broda}}, \bibinfo {author} {\bibfnamefont
  {M.}~\bibnamefont {Gębski}}, \bibinfo {author} {\bibfnamefont
  {J.}~\bibnamefont {Muszalski}}, \bibinfo {author} {\bibfnamefont
  {D.}~\bibnamefont {Pudiš}}, \bibinfo {author} {\bibfnamefont
  {J.}~\bibnamefont {Suffczyński}},\ and\ \bibinfo {author} {\bibfnamefont
  {T.}~\bibnamefont {Czyszanowski}},\ }\bibfield  {title} {\bibinfo {title}
  {Concept of inverted refractive-index-contrast grating mirror and exemplary
  fabrication by 3d laser micro-printing},\ }\href
  {https://doi.org/10.1515/nanoph-2023-0283} {\bibfield  {journal} {\bibinfo
  {journal} {Nanophotonics}\ }\textbf {\bibinfo {volume} {12}},\ \bibinfo
  {pages} {3579} (\bibinfo {year} {2023})}\BibitemShut {NoStop}%
\bibitem [{\citenamefont {Duong}\ \emph {et~al.}(2017)\citenamefont {Duong},
  \citenamefont {Yun},\ and\ \citenamefont {Lee}}]{Duong2017ACSN}%
  \BibitemOpen
  \bibfield  {author} {\bibinfo {author} {\bibfnamefont {D.~L.}\ \bibnamefont
  {Duong}}, \bibinfo {author} {\bibfnamefont {S.~J.}\ \bibnamefont {Yun}},\
  and\ \bibinfo {author} {\bibfnamefont {Y.~H.}\ \bibnamefont {Lee}},\
  }\bibfield  {title} {\bibinfo {title} {van der waals layered materials:
  opportunities and challenges},\ }\href@noop {} {\bibfield  {journal}
  {\bibinfo  {journal} {ACS nano}\ }\textbf {\bibinfo {volume} {11}},\ \bibinfo
  {pages} {11803} (\bibinfo {year} {2017})}\BibitemShut {NoStop}%
\bibitem [{\citenamefont {He}\ \emph {et~al.}(2019)\citenamefont {He},
  \citenamefont {Wang}, \citenamefont {Zhong}, \citenamefont {Fang},
  \citenamefont {Wang},\ and\ \citenamefont {Hu}}]{Ting:AMT2019}%
  \BibitemOpen
  \bibfield  {author} {\bibinfo {author} {\bibfnamefont {T.}~\bibnamefont
  {He}}, \bibinfo {author} {\bibfnamefont {Z.}~\bibnamefont {Wang}}, \bibinfo
  {author} {\bibfnamefont {F.}~\bibnamefont {Zhong}}, \bibinfo {author}
  {\bibfnamefont {H.}~\bibnamefont {Fang}}, \bibinfo {author} {\bibfnamefont
  {P.}~\bibnamefont {Wang}},\ and\ \bibinfo {author} {\bibfnamefont
  {W.}~\bibnamefont {Hu}},\ }\bibfield  {title} {\bibinfo {title} {Etching
  techniques in 2d materials},\ }\href
  {https://doi.org/https://doi.org/10.1002/admt.201900064} {\bibfield
  {journal} {\bibinfo  {journal} {Advanced Materials Technologies}\ }\textbf
  {\bibinfo {volume} {4}},\ \bibinfo {pages} {1900064} (\bibinfo {year}
  {2019})},\ \Eprint
  {https://arxiv.org/abs/https://onlinelibrary.wiley.com/doi/pdf/10.1002/admt.201900064}
  {https://onlinelibrary.wiley.com/doi/pdf/10.1002/admt.201900064} \BibitemShut
  {NoStop}%
\bibitem [{\citenamefont {Singh}\ \emph {et~al.}(2020)\citenamefont {Singh},
  \citenamefont {Jo}, \citenamefont {Li}, \citenamefont {Wu}, \citenamefont
  {Li},\ and\ \citenamefont {Jaramillo}}]{Singh:ACSPhotonics2020}%
  \BibitemOpen
  \bibfield  {author} {\bibinfo {author} {\bibfnamefont {A.}~\bibnamefont
  {Singh}}, \bibinfo {author} {\bibfnamefont {S.~S.}\ \bibnamefont {Jo}},
  \bibinfo {author} {\bibfnamefont {Y.}~\bibnamefont {Li}}, \bibinfo {author}
  {\bibfnamefont {C.}~\bibnamefont {Wu}}, \bibinfo {author} {\bibfnamefont
  {M.}~\bibnamefont {Li}},\ and\ \bibinfo {author} {\bibfnamefont
  {R.}~\bibnamefont {Jaramillo}},\ }\bibfield  {title} {\bibinfo {title}
  {{Refractive Uses of Layered and Two-Dimensional Materials for Integrated
  Photonics}},\ }\href {https://doi.org/10.1021/acsphotonics.0c00915}
  {\bibfield  {journal} {\bibinfo  {journal} {ACS Photonics}\ }\textbf
  {\bibinfo {volume} {7}},\ \bibinfo {pages} {3270} (\bibinfo {year}
  {2020})}\BibitemShut {NoStop}%
\bibitem [{\citenamefont {Lin}\ \emph {et~al.}(2020)\citenamefont {Lin},
  \citenamefont {Xu}, \citenamefont {Cao}, \citenamefont {Zhang}, \citenamefont
  {Zhou}, \citenamefont {Wang}, \citenamefont {Wan}, \citenamefont {Liu},
  \citenamefont {Loh}, \citenamefont {Qiu}, \citenamefont {Bao},\ and\
  \citenamefont {Jia}}]{Lin2020LSA}%
  \BibitemOpen
  \bibfield  {author} {\bibinfo {author} {\bibfnamefont {H.}~\bibnamefont
  {Lin}}, \bibinfo {author} {\bibfnamefont {Z.~Q.}\ \bibnamefont {Xu}},
  \bibinfo {author} {\bibfnamefont {G.}~\bibnamefont {Cao}}, \bibinfo {author}
  {\bibfnamefont {Y.}~\bibnamefont {Zhang}}, \bibinfo {author} {\bibfnamefont
  {J.}~\bibnamefont {Zhou}}, \bibinfo {author} {\bibfnamefont {Z.}~\bibnamefont
  {Wang}}, \bibinfo {author} {\bibfnamefont {Z.}~\bibnamefont {Wan}}, \bibinfo
  {author} {\bibfnamefont {Z.}~\bibnamefont {Liu}}, \bibinfo {author}
  {\bibfnamefont {K.~P.}\ \bibnamefont {Loh}}, \bibinfo {author} {\bibfnamefont
  {C.~W.}\ \bibnamefont {Qiu}}, \bibinfo {author} {\bibfnamefont
  {Q.}~\bibnamefont {Bao}},\ and\ \bibinfo {author} {\bibfnamefont
  {B.}~\bibnamefont {Jia}},\ }\bibfield  {title} {\bibinfo {title}
  {Diffraction-limited imaging with monolayer 2d material-based ultrathin flat
  lenses},\ }\href {https://doi.org/10.1038/s41377-020-00374-9} {\bibfield
  {journal} {\bibinfo  {journal} {Light: Science and Applications}\ }\textbf
  {\bibinfo {volume} {9}},\ \bibinfo {pages} {137} (\bibinfo {year}
  {2020})}\BibitemShut {NoStop}%
\bibitem [{\citenamefont {Zhang}\ \emph {et~al.}(2019)\citenamefont {Zhang},
  \citenamefont {De-Eknamkul}, \citenamefont {Gu}, \citenamefont {Boehmke},
  \citenamefont {Menon}, \citenamefont {Khurgin},\ and\ \citenamefont
  {Cubukcu}}]{Zhang2019NN}%
  \BibitemOpen
  \bibfield  {author} {\bibinfo {author} {\bibfnamefont {X.}~\bibnamefont
  {Zhang}}, \bibinfo {author} {\bibfnamefont {C.}~\bibnamefont {De-Eknamkul}},
  \bibinfo {author} {\bibfnamefont {J.}~\bibnamefont {Gu}}, \bibinfo {author}
  {\bibfnamefont {A.~L.}\ \bibnamefont {Boehmke}}, \bibinfo {author}
  {\bibfnamefont {V.~M.}\ \bibnamefont {Menon}}, \bibinfo {author}
  {\bibfnamefont {J.}~\bibnamefont {Khurgin}},\ and\ \bibinfo {author}
  {\bibfnamefont {E.}~\bibnamefont {Cubukcu}},\ }\bibfield  {title} {\bibinfo
  {title} {Guiding of visible photons at the ångström thickness limit},\
  }\href {https://doi.org/10.1038/s41565-019-0519-6} {\bibfield  {journal}
  {\bibinfo  {journal} {Nature Nanotechnology}\ }\textbf {\bibinfo {volume}
  {14}},\ \bibinfo {pages} {844} (\bibinfo {year} {2019})}\BibitemShut
  {NoStop}%
\bibitem [{\citenamefont {Guarneri}\ \emph {et~al.}(2024)\citenamefont
  {Guarneri}, \citenamefont {Li}, \citenamefont {Bauer}, \citenamefont {Song},
  \citenamefont {Saunders}, \citenamefont {Liu}, \citenamefont {Brongersma},\
  and\ \citenamefont {van~de Groep}}]{Guarneri2024NL}%
  \BibitemOpen
  \bibfield  {author} {\bibinfo {author} {\bibfnamefont {L.}~\bibnamefont
  {Guarneri}}, \bibinfo {author} {\bibfnamefont {Q.}~\bibnamefont {Li}},
  \bibinfo {author} {\bibfnamefont {T.}~\bibnamefont {Bauer}}, \bibinfo
  {author} {\bibfnamefont {J.-H.}\ \bibnamefont {Song}}, \bibinfo {author}
  {\bibfnamefont {A.~P.}\ \bibnamefont {Saunders}}, \bibinfo {author}
  {\bibfnamefont {F.}~\bibnamefont {Liu}}, \bibinfo {author} {\bibfnamefont
  {M.~L.}\ \bibnamefont {Brongersma}},\ and\ \bibinfo {author} {\bibfnamefont
  {J.}~\bibnamefont {van~de Groep}},\ }\bibfield  {title} {\bibinfo {title}
  {Temperature-dependent excitonic light manipulation with atomically thin
  optical elements},\ }\href {https://doi.org/10.1021/acs.nanolett.4c00694}
  {\bibfield  {journal} {\bibinfo  {journal} {Nano Letters}\ }\textbf {\bibinfo
  {volume} {24}},\ \bibinfo {pages} {6240} (\bibinfo {year}
  {2024})}\BibitemShut {NoStop}%
\bibitem [{\citenamefont {Zong}\ \emph {et~al.}(2021)\citenamefont {Zong},
  \citenamefont {Li},\ and\ \citenamefont {Liu}}]{Zong2021OL}%
  \BibitemOpen
  \bibfield  {author} {\bibinfo {author} {\bibfnamefont {X.}~\bibnamefont
  {Zong}}, \bibinfo {author} {\bibfnamefont {L.}~\bibnamefont {Li}},\ and\
  \bibinfo {author} {\bibfnamefont {Y.}~\bibnamefont {Liu}},\ }\bibfield
  {title} {\bibinfo {title} {Photonic bound states in the continuum in
  nanostructured transition metal dichalcogenides for strong photon–exciton
  coupling},\ }\href {https://doi.org/10.1364/ol.446950} {\bibfield  {journal}
  {\bibinfo  {journal} {Optics Letters}\ }\textbf {\bibinfo {volume} {46}},\
  \bibinfo {pages} {6095} (\bibinfo {year} {2021})}\BibitemShut {NoStop}%
\bibitem [{\citenamefont {Qin}\ \emph {et~al.}(2023)\citenamefont {Qin},
  \citenamefont {Duan}, \citenamefont {Xiao}, \citenamefont {Liu},
  \citenamefont {Yu}, \citenamefont {Wang},\ and\ \citenamefont
  {Liao}}]{Qin2023PRB}%
  \BibitemOpen
  \bibfield  {author} {\bibinfo {author} {\bibfnamefont {M.}~\bibnamefont
  {Qin}}, \bibinfo {author} {\bibfnamefont {J.}~\bibnamefont {Duan}}, \bibinfo
  {author} {\bibfnamefont {S.}~\bibnamefont {Xiao}}, \bibinfo {author}
  {\bibfnamefont {W.}~\bibnamefont {Liu}}, \bibinfo {author} {\bibfnamefont
  {T.}~\bibnamefont {Yu}}, \bibinfo {author} {\bibfnamefont {T.}~\bibnamefont
  {Wang}},\ and\ \bibinfo {author} {\bibfnamefont {Q.}~\bibnamefont {Liao}},\
  }\bibfield  {title} {\bibinfo {title} {Strong coupling between excitons and
  quasibound states in the continuum in bulk transition metal
  dichalcogenides},\ }\bibfield  {journal} {\bibinfo  {journal} {Physical
  Review B}\ }\textbf {\bibinfo {volume} {107}},\ \href
  {https://doi.org/10.1103/PhysRevB.107.045417} {10.1103/PhysRevB.107.045417}
  (\bibinfo {year} {2023})\BibitemShut {NoStop}%
\bibitem [{\citenamefont {Isoniemi}\ \emph {et~al.}(2024)\citenamefont
  {Isoniemi}, \citenamefont {Bouteyre}, \citenamefont {Hu}, \citenamefont
  {Benimetskiy}, \citenamefont {Wang}, \citenamefont {Skolnick}, \citenamefont
  {Krizhanovskii},\ and\ \citenamefont {Tartakovskii}}]{Isoniemi:ACSNano2024}%
  \BibitemOpen
  \bibfield  {author} {\bibinfo {author} {\bibfnamefont {T.}~\bibnamefont
  {Isoniemi}}, \bibinfo {author} {\bibfnamefont {P.}~\bibnamefont {Bouteyre}},
  \bibinfo {author} {\bibfnamefont {X.}~\bibnamefont {Hu}}, \bibinfo {author}
  {\bibfnamefont {F.}~\bibnamefont {Benimetskiy}}, \bibinfo {author}
  {\bibfnamefont {Y.}~\bibnamefont {Wang}}, \bibinfo {author} {\bibfnamefont
  {M.~S.}\ \bibnamefont {Skolnick}}, \bibinfo {author} {\bibfnamefont {D.~N.}\
  \bibnamefont {Krizhanovskii}},\ and\ \bibinfo {author} {\bibfnamefont
  {A.~I.}\ \bibnamefont {Tartakovskii}},\ }\bibfield  {title} {\bibinfo {title}
  {Realization of {Z2} topological photonic insulators made from multilayer
  transition metal dichalcogenides},\ }\href
  {https://doi.org/10.1021/acsnano.4c09295} {\bibfield  {journal} {\bibinfo
  {journal} {ACS Nano}\ }\textbf {\bibinfo {volume} {18}},\ \bibinfo {pages}
  {32547} (\bibinfo {year} {2024})}\BibitemShut {NoStop}%
\bibitem [{\citenamefont {Maggiolini}\ \emph {et~al.}(2023)\citenamefont
  {Maggiolini}, \citenamefont {Polimeno}, \citenamefont {Todisco},
  \citenamefont {Di~Renzo}, \citenamefont {Han}, \citenamefont {De~Giorgi},
  \citenamefont {Ardizzone}, \citenamefont {Schneider}, \citenamefont
  {Mastria}, \citenamefont {Cannavale}, \citenamefont {Pugliese}, \citenamefont
  {De~Marco}, \citenamefont {Rizzo}, \citenamefont {Maiorano}, \citenamefont
  {Gigli}, \citenamefont {Gerace}, \citenamefont {Sanvitto},\ and\
  \citenamefont {Ballarini}}]{Maggiolini:NatureMat2023}%
  \BibitemOpen
  \bibfield  {author} {\bibinfo {author} {\bibfnamefont {E.}~\bibnamefont
  {Maggiolini}}, \bibinfo {author} {\bibfnamefont {L.}~\bibnamefont
  {Polimeno}}, \bibinfo {author} {\bibfnamefont {F.}~\bibnamefont {Todisco}},
  \bibinfo {author} {\bibfnamefont {A.}~\bibnamefont {Di~Renzo}}, \bibinfo
  {author} {\bibfnamefont {B.}~\bibnamefont {Han}}, \bibinfo {author}
  {\bibfnamefont {M.}~\bibnamefont {De~Giorgi}}, \bibinfo {author}
  {\bibfnamefont {V.}~\bibnamefont {Ardizzone}}, \bibinfo {author}
  {\bibfnamefont {C.}~\bibnamefont {Schneider}}, \bibinfo {author}
  {\bibfnamefont {R.}~\bibnamefont {Mastria}}, \bibinfo {author} {\bibfnamefont
  {A.}~\bibnamefont {Cannavale}}, \bibinfo {author} {\bibfnamefont
  {M.}~\bibnamefont {Pugliese}}, \bibinfo {author} {\bibfnamefont
  {L.}~\bibnamefont {De~Marco}}, \bibinfo {author} {\bibfnamefont
  {A.}~\bibnamefont {Rizzo}}, \bibinfo {author} {\bibfnamefont
  {V.}~\bibnamefont {Maiorano}}, \bibinfo {author} {\bibfnamefont
  {G.}~\bibnamefont {Gigli}}, \bibinfo {author} {\bibfnamefont
  {D.}~\bibnamefont {Gerace}}, \bibinfo {author} {\bibfnamefont
  {D.}~\bibnamefont {Sanvitto}},\ and\ \bibinfo {author} {\bibfnamefont
  {D.}~\bibnamefont {Ballarini}},\ }\bibfield  {title} {\bibinfo {title}
  {{Strongly enhanced light{\textendash}matter coupling of monolayer WS2 from a
  bound state in the continuum}},\ }\href
  {https://doi.org/10.1038/s41563-023-01562-9} {\bibfield  {journal} {\bibinfo
  {journal} {Nat. Mater.}\ }\textbf {\bibinfo {volume} {22}},\ \bibinfo {pages}
  {964} (\bibinfo {year} {2023})}\BibitemShut {NoStop}%
\bibitem [{\citenamefont {Ning}\ \emph {et~al.}(2020)\citenamefont {Ning},
  \citenamefont {Li}, \citenamefont {Zhao}, \citenamefont {Yin}, \citenamefont
  {Huo}, \citenamefont {Zhao},\ and\ \citenamefont {Yue}}]{Ning2020OE}%
  \BibitemOpen
  \bibfield  {author} {\bibinfo {author} {\bibfnamefont {T.}~\bibnamefont
  {Ning}}, \bibinfo {author} {\bibfnamefont {X.}~\bibnamefont {Li}}, \bibinfo
  {author} {\bibfnamefont {Y.}~\bibnamefont {Zhao}}, \bibinfo {author}
  {\bibfnamefont {L.}~\bibnamefont {Yin}}, \bibinfo {author} {\bibfnamefont
  {Y.}~\bibnamefont {Huo}}, \bibinfo {author} {\bibfnamefont {L.}~\bibnamefont
  {Zhao}},\ and\ \bibinfo {author} {\bibfnamefont {Q.}~\bibnamefont {Yue}},\
  }\bibfield  {title} {\bibinfo {title} {Giant enhancement of harmonic
  generation in all-dielectric resonant waveguide gratings of quasi-bound
  states in the continuum},\ }\href@noop {} {\bibfield  {journal} {\bibinfo
  {journal} {Optics Express}\ }\textbf {\bibinfo {volume} {28}},\ \bibinfo
  {pages} {34024} (\bibinfo {year} {2020})}\BibitemShut {NoStop}%
\bibitem [{\citenamefont {Shcherbakov}\ \emph {et~al.}(2021)\citenamefont
  {Shcherbakov}, \citenamefont {Zhang}, \citenamefont {Tripepi}, \citenamefont
  {Sartorello}, \citenamefont {Talisa}, \citenamefont {AlShafey}, \citenamefont
  {Fan}, \citenamefont {Twardowski}, \citenamefont {Krivitsky}, \citenamefont
  {Kuznetsov} \emph {et~al.}}]{Shcherbakov2021NC}%
  \BibitemOpen
  \bibfield  {author} {\bibinfo {author} {\bibfnamefont {M.~R.}\ \bibnamefont
  {Shcherbakov}}, \bibinfo {author} {\bibfnamefont {H.}~\bibnamefont {Zhang}},
  \bibinfo {author} {\bibfnamefont {M.}~\bibnamefont {Tripepi}}, \bibinfo
  {author} {\bibfnamefont {G.}~\bibnamefont {Sartorello}}, \bibinfo {author}
  {\bibfnamefont {N.}~\bibnamefont {Talisa}}, \bibinfo {author} {\bibfnamefont
  {A.}~\bibnamefont {AlShafey}}, \bibinfo {author} {\bibfnamefont
  {Z.}~\bibnamefont {Fan}}, \bibinfo {author} {\bibfnamefont {J.}~\bibnamefont
  {Twardowski}}, \bibinfo {author} {\bibfnamefont {L.~A.}\ \bibnamefont
  {Krivitsky}}, \bibinfo {author} {\bibfnamefont {A.~I.}\ \bibnamefont
  {Kuznetsov}}, \emph {et~al.},\ }\bibfield  {title} {\bibinfo {title}
  {Generation of even and odd high harmonics in resonant metasurfaces using
  single and multiple ultra-intense laser pulses},\ }\href@noop {} {\bibfield
  {journal} {\bibinfo  {journal} {Nature communications}\ }\textbf {\bibinfo
  {volume} {12}},\ \bibinfo {pages} {4185} (\bibinfo {year}
  {2021})}\BibitemShut {NoStop}%
\bibitem [{\citenamefont {Qiu}\ \emph {et~al.}(2023)\citenamefont {Qiu},
  \citenamefont {Yan}, \citenamefont {Li}, \citenamefont {Zhang},\ and\
  \citenamefont {Li}}]{Qiu2023OC}%
  \BibitemOpen
  \bibfield  {author} {\bibinfo {author} {\bibfnamefont {Y.}~\bibnamefont
  {Qiu}}, \bibinfo {author} {\bibfnamefont {D.}~\bibnamefont {Yan}}, \bibinfo
  {author} {\bibfnamefont {X.}~\bibnamefont {Li}}, \bibinfo {author}
  {\bibfnamefont {L.}~\bibnamefont {Zhang}},\ and\ \bibinfo {author}
  {\bibfnamefont {J.}~\bibnamefont {Li}},\ }\bibfield  {title} {\bibinfo
  {title} {Highly efficient second harmonic generation assisted by the
  quasi-bound states in the continuum from algaas meta-gratings},\ }\href@noop
  {} {\bibfield  {journal} {\bibinfo  {journal} {Optics Communications}\
  }\textbf {\bibinfo {volume} {546}},\ \bibinfo {pages} {129772} (\bibinfo
  {year} {2023})}\BibitemShut {NoStop}%
\bibitem [{\citenamefont {Wang}\ \emph {et~al.}(2024)\citenamefont {Wang},
  \citenamefont {You},\ and\ \citenamefont {Panoiu}}]{Wang:Nanophotonics2024}%
  \BibitemOpen
  \bibfield  {author} {\bibinfo {author} {\bibfnamefont {J.~T.}\ \bibnamefont
  {Wang}}, \bibinfo {author} {\bibfnamefont {J.~W.}\ \bibnamefont {You}},\ and\
  \bibinfo {author} {\bibfnamefont {N.~C.}\ \bibnamefont {Panoiu}},\ }\bibfield
   {title} {\bibinfo {title} {Giant second-harmonic generation in monolayer
  mos2 boosted by dual bound states in the continuum},\ }\href
  {https://doi.org/doi:10.1515/nanoph-2024-0273} {\bibfield  {journal}
  {\bibinfo  {journal} {Nanophotonics}\ }\textbf {\bibinfo {volume} {13}},\
  \bibinfo {pages} {3437} (\bibinfo {year} {2024})}\BibitemShut {NoStop}%
\bibitem [{\citenamefont {Lee}\ \emph {et~al.}(2025)\citenamefont {Lee},
  \citenamefont {Jeong}, \citenamefont {Jang}, \citenamefont {Kim},
  \citenamefont {Mun}, \citenamefont {Gong}, \citenamefont {Fang},
  \citenamefont {Yang}, \citenamefont {Chae}, \citenamefont {Kim},\ and\
  \citenamefont {Rho}}]{Lee:NanoLett2025}%
  \BibitemOpen
  \bibfield  {author} {\bibinfo {author} {\bibfnamefont {J.}~\bibnamefont
  {Lee}}, \bibinfo {author} {\bibfnamefont {M.}~\bibnamefont {Jeong}}, \bibinfo
  {author} {\bibfnamefont {J.}~\bibnamefont {Jang}}, \bibinfo {author}
  {\bibfnamefont {J.}~\bibnamefont {Kim}}, \bibinfo {author} {\bibfnamefont
  {J.}~\bibnamefont {Mun}}, \bibinfo {author} {\bibfnamefont {X.}~\bibnamefont
  {Gong}}, \bibinfo {author} {\bibfnamefont {R.}~\bibnamefont {Fang}}, \bibinfo
  {author} {\bibfnamefont {Y.}~\bibnamefont {Yang}}, \bibinfo {author}
  {\bibfnamefont {S.~H.}\ \bibnamefont {Chae}}, \bibinfo {author}
  {\bibfnamefont {S.}~\bibnamefont {Kim}},\ and\ \bibinfo {author}
  {\bibfnamefont {J.}~\bibnamefont {Rho}},\ }\bibfield  {title} {\bibinfo
  {title} {{Bound-States-in-the-Continuum-Induced Directional Photoluminescence
  with Polarization Singularity in WS2 Monolayers}},\ }\bibfield  {journal}
  {\bibinfo  {journal} {Nano Lett.}\ }\href
  {https://doi.org/10.1021/acs.nanolett.4c05544} {10.1021/acs.nanolett.4c05544}
  (\bibinfo {year} {2025})\BibitemShut {NoStop}%
\bibitem [{\citenamefont {Pacuski}\ \emph {et~al.}(2020)\citenamefont
  {Pacuski}, \citenamefont {Grzeszczyk}, \citenamefont {Nogajewski},
  \citenamefont {Bogucki}, \citenamefont {Oreszczuk}, \citenamefont {Kucharek},
  \citenamefont {Po{\l}czyńska}, \citenamefont {Seredyński}, \citenamefont
  {Rodek}, \citenamefont {Bożek} \emph {et~al.}}]{Pacuski2020NL}%
  \BibitemOpen
  \bibfield  {author} {\bibinfo {author} {\bibfnamefont {W.}~\bibnamefont
  {Pacuski}}, \bibinfo {author} {\bibfnamefont {M.}~\bibnamefont {Grzeszczyk}},
  \bibinfo {author} {\bibfnamefont {K.}~\bibnamefont {Nogajewski}}, \bibinfo
  {author} {\bibfnamefont {A.}~\bibnamefont {Bogucki}}, \bibinfo {author}
  {\bibfnamefont {K.}~\bibnamefont {Oreszczuk}}, \bibinfo {author}
  {\bibfnamefont {J.}~\bibnamefont {Kucharek}}, \bibinfo {author}
  {\bibfnamefont {K.~E.}\ \bibnamefont {Po{\l}czyńska}}, \bibinfo {author}
  {\bibfnamefont {B.}~\bibnamefont {Seredyński}}, \bibinfo {author}
  {\bibfnamefont {A.}~\bibnamefont {Rodek}}, \bibinfo {author} {\bibfnamefont
  {R.}~\bibnamefont {Bożek}}, \emph {et~al.},\ }\bibfield  {title} {\bibinfo
  {title} {Narrow excitonic lines and large-scale homogeneity of
  transition-metal dichalcogenide monolayers grown by molecular beam epitaxy on
  hexagonal boron nitride},\ }\href@noop {} {\bibfield  {journal} {\bibinfo
  {journal} {Nano letters}\ }\textbf {\bibinfo {volume} {20}},\ \bibinfo
  {pages} {3058} (\bibinfo {year} {2020})}\BibitemShut {NoStop}%
\bibitem [{\citenamefont {Kong}\ \emph {et~al.}(2013)\citenamefont {Kong},
  \citenamefont {Wang}, \citenamefont {Cha}, \citenamefont {Pasta},
  \citenamefont {Koski}, \citenamefont {Yao},\ and\ \citenamefont
  {Cui}}]{Kong2013NL}%
  \BibitemOpen
  \bibfield  {author} {\bibinfo {author} {\bibfnamefont {D.}~\bibnamefont
  {Kong}}, \bibinfo {author} {\bibfnamefont {H.}~\bibnamefont {Wang}}, \bibinfo
  {author} {\bibfnamefont {J.~J.}\ \bibnamefont {Cha}}, \bibinfo {author}
  {\bibfnamefont {M.}~\bibnamefont {Pasta}}, \bibinfo {author} {\bibfnamefont
  {K.~J.}\ \bibnamefont {Koski}}, \bibinfo {author} {\bibfnamefont
  {J.}~\bibnamefont {Yao}},\ and\ \bibinfo {author} {\bibfnamefont
  {Y.}~\bibnamefont {Cui}},\ }\bibfield  {title} {\bibinfo {title} {{Synthesis
  of MoS2 and MoSe2 Films with Vertically Aligned Layers}},\ }\href
  {https://doi.org/10.1021/nl400258t} {\bibfield  {journal} {\bibinfo
  {journal} {Nano Lett.}\ }\textbf {\bibinfo {volume} {13}},\ \bibinfo {pages}
  {1341} (\bibinfo {year} {2013})}\BibitemShut {NoStop}%
\bibitem [{\citenamefont {Li}\ \emph {et~al.}(2016)\citenamefont {Li},
  \citenamefont {Wu}, \citenamefont {Yuan},\ and\ \citenamefont
  {Qian}}]{Li2016SR}%
  \BibitemOpen
  \bibfield  {author} {\bibinfo {author} {\bibfnamefont {H.}~\bibnamefont
  {Li}}, \bibinfo {author} {\bibfnamefont {H.}~\bibnamefont {Wu}}, \bibinfo
  {author} {\bibfnamefont {S.}~\bibnamefont {Yuan}},\ and\ \bibinfo {author}
  {\bibfnamefont {H.}~\bibnamefont {Qian}},\ }\bibfield  {title} {\bibinfo
  {title} {{Synthesis and characterization of vertically standing MoS2
  nanosheets}},\ }\href {https://doi.org/10.1038/srep21171} {\bibfield
  {journal} {\bibinfo  {journal} {Sci. Rep.}\ }\textbf {\bibinfo {volume}
  {6}},\ \bibinfo {pages} {1} (\bibinfo {year} {2016})}\BibitemShut {NoStop}%
\bibitem [{\citenamefont {Bouteyre}\ \emph {et~al.}(2022)\citenamefont
  {Bouteyre}, \citenamefont {Nguyen}, \citenamefont {Gachon}, \citenamefont
  {Benyattou}, \citenamefont {Letartre}, \citenamefont {Viktorovitch},
  \citenamefont {Callard}, \citenamefont {Ferrier},\ and\ \citenamefont
  {Nguyen}}]{Nguyen:Vortex}%
  \BibitemOpen
  \bibfield  {author} {\bibinfo {author} {\bibfnamefont {P.}~\bibnamefont
  {Bouteyre}}, \bibinfo {author} {\bibfnamefont {D.~X.}\ \bibnamefont
  {Nguyen}}, \bibinfo {author} {\bibfnamefont {G.}~\bibnamefont {Gachon}},
  \bibinfo {author} {\bibfnamefont {T.}~\bibnamefont {Benyattou}}, \bibinfo
  {author} {\bibfnamefont {X.}~\bibnamefont {Letartre}}, \bibinfo {author}
  {\bibfnamefont {P.}~\bibnamefont {Viktorovitch}}, \bibinfo {author}
  {\bibfnamefont {S.}~\bibnamefont {Callard}}, \bibinfo {author} {\bibfnamefont
  {L.}~\bibnamefont {Ferrier}},\ and\ \bibinfo {author} {\bibfnamefont {H.~S.}\
  \bibnamefont {Nguyen}},\ }\bibfield  {title} {\bibinfo {title} {Non-hermitian
  topological invariant of photonic band structures undergoing inversion},\
  }\href@noop {} {\bibfield  {journal} {\bibinfo  {journal} {arXiv preprint
  arXiv:2211.09884}\ } (\bibinfo {year} {2022})}\BibitemShut {NoStop}%
\bibitem [{\citenamefont {Zhen}\ \emph {et~al.}(2014)\citenamefont {Zhen},
  \citenamefont {Hsu}, \citenamefont {Lu}, \citenamefont {Stone},\ and\
  \citenamefont {Solja\ifmmode \check{c}\else
  \v{c}\fi{}i\ifmmode~\acute{c}\else \'{c}\fi{}}}]{Zhen:PRL2013}%
  \BibitemOpen
  \bibfield  {author} {\bibinfo {author} {\bibfnamefont {B.}~\bibnamefont
  {Zhen}}, \bibinfo {author} {\bibfnamefont {C.~W.}\ \bibnamefont {Hsu}},
  \bibinfo {author} {\bibfnamefont {L.}~\bibnamefont {Lu}}, \bibinfo {author}
  {\bibfnamefont {A.~D.}\ \bibnamefont {Stone}},\ and\ \bibinfo {author}
  {\bibfnamefont {M.}~\bibnamefont {Solja\ifmmode \check{c}\else
  \v{c}\fi{}i\ifmmode~\acute{c}\else \'{c}\fi{}}},\ }\bibfield  {title}
  {\bibinfo {title} {Topological nature of optical bound states in the
  continuum},\ }\href {https://doi.org/10.1103/PhysRevLett.113.257401}
  {\bibfield  {journal} {\bibinfo  {journal} {Phys. Rev. Lett.}\ }\textbf
  {\bibinfo {volume} {113}},\ \bibinfo {pages} {257401} (\bibinfo {year}
  {2014})}\BibitemShut {NoStop}%
\bibitem [{\citenamefont {Khan}\ \emph {et~al.}(2020)\citenamefont {Khan},
  \citenamefont {Liu}, \citenamefont {Zhang}, \citenamefont {Zhu},
  \citenamefont {He}, \citenamefont {Zhang}, \citenamefont {Lü},\ and\
  \citenamefont {Lu}}]{Khan2020}%
  \BibitemOpen
  \bibfield  {author} {\bibinfo {author} {\bibfnamefont {A.~R.}\ \bibnamefont
  {Khan}}, \bibinfo {author} {\bibfnamefont {B.}~\bibnamefont {Liu}}, \bibinfo
  {author} {\bibfnamefont {L.}~\bibnamefont {Zhang}}, \bibinfo {author}
  {\bibfnamefont {Y.}~\bibnamefont {Zhu}}, \bibinfo {author} {\bibfnamefont
  {X.}~\bibnamefont {He}}, \bibinfo {author} {\bibfnamefont {L.}~\bibnamefont
  {Zhang}}, \bibinfo {author} {\bibfnamefont {T.}~\bibnamefont {Lü}},\ and\
  \bibinfo {author} {\bibfnamefont {Y.}~\bibnamefont {Lu}},\ }\bibfield
  {title} {\bibinfo {title} {Extraordinary temperature dependent second
  harmonic generation in atomically thin layers of transition-metal
  dichalcogenides},\ }\href
  {https://doi.org/https://doi.org/10.1002/adom.202000441} {\bibfield
  {journal} {\bibinfo  {journal} {Advanced Optical Materials}\ }\textbf
  {\bibinfo {volume} {8}},\ \bibinfo {pages} {2000441} (\bibinfo {year}
  {2020})},\ \Eprint
  {https://arxiv.org/abs/https://onlinelibrary.wiley.com/doi/pdf/10.1002/adom.202000441}
  {https://onlinelibrary.wiley.com/doi/pdf/10.1002/adom.202000441} \BibitemShut
  {NoStop}%
\bibitem [{\citenamefont {Dems}\ \emph {et~al.}(2005)\citenamefont {Dems},
  \citenamefont {Kotynski},\ and\ \citenamefont {Panajotov}}]{Dems2007PhD}%
  \BibitemOpen
  \bibfield  {author} {\bibinfo {author} {\bibfnamefont {M.}~\bibnamefont
  {Dems}}, \bibinfo {author} {\bibfnamefont {R.}~\bibnamefont {Kotynski}},\
  and\ \bibinfo {author} {\bibfnamefont {K.}~\bibnamefont {Panajotov}},\
  }\bibfield  {title} {\bibinfo {title} {Planewave admittance method—a novel
  approach for determining the electromagnetic modes in photonic structures},\
  }\href@noop {} {\bibfield  {journal} {\bibinfo  {journal} {Optics Express}\
  }\textbf {\bibinfo {volume} {13}},\ \bibinfo {pages} {3196} (\bibinfo {year}
  {2005})}\BibitemShut {NoStop}%
\bibitem [{\citenamefont {Dems}(2011)}]{Dems2011OER}%
  \BibitemOpen
  \bibfield  {author} {\bibinfo {author} {\bibfnamefont {M.}~\bibnamefont
  {Dems}},\ }\bibfield  {title} {\bibinfo {title} {Modelling of high-contrast
  grating mirrors. the impact of imperfections on their performance in
  vcsels},\ }\href {https://doi.org/https://doi.org/10.2478/s11772-011-0027-1}
  {\bibfield  {journal} {\bibinfo  {journal} {Opto-Electronics Review}\
  }\textbf {\bibinfo {volume} {19}},\ \bibinfo {pages} {340} (\bibinfo {year}
  {2011})}\BibitemShut {NoStop}%
\bibitem [{\citenamefont {Chartier}(2005)}]{Chartier_Optics}%
  \BibitemOpen
  \bibfield  {author} {\bibinfo {author} {\bibfnamefont {G.}~\bibnamefont
  {Chartier}},\ }\href@noop {} {\emph {\bibinfo {title} {Introduction to
  Optics}}}\ (\bibinfo  {publisher} {Springer Science+Business Media, Inc},\
  \bibinfo {year} {2005})\ p.\ \bibinfo {pages} {284}\BibitemShut {NoStop}%
\bibitem [{\citenamefont {Novotny}\ and\ \citenamefont
  {Hecht}(2012)}]{Novotny2012}%
  \BibitemOpen
  \bibfield  {author} {\bibinfo {author} {\bibfnamefont {L.}~\bibnamefont
  {Novotny}}\ and\ \bibinfo {author} {\bibfnamefont {B.}~\bibnamefont
  {Hecht}},\ }\href@noop {} {\emph {\bibinfo {title} {Principles of
  Nano-Optics}}}\ (\bibinfo  {publisher} {Cambridge University Press},\
  \bibinfo {year} {2012})\BibitemShut {NoStop}%
\bibitem [{\citenamefont {Pruszyńska-Karbownik}\ \emph
  {et~al.}(2014)\citenamefont {Pruszyńska-Karbownik}, \citenamefont
  {Regiński}, \citenamefont {Karbownik},\ and\ \citenamefont
  {Mroziewicz}}]{EPK2014OQE}%
  \BibitemOpen
  \bibfield  {author} {\bibinfo {author} {\bibfnamefont {E.}~\bibnamefont
  {Pruszyńska-Karbownik}}, \bibinfo {author} {\bibfnamefont {K.}~\bibnamefont
  {Regiński}}, \bibinfo {author} {\bibfnamefont {P.}~\bibnamefont
  {Karbownik}},\ and\ \bibinfo {author} {\bibfnamefont {B.}~\bibnamefont
  {Mroziewicz}},\ }\bibfield  {title} {\bibinfo {title} {Intra-pulse beam
  steering in a mid-infrared quantum cascade laser},\ }\href
  {https://doi.org/10.1007/s11082-014-0006-0} {\bibfield  {journal} {\bibinfo
  {journal} {Optical and Quantum Electronics}\ }\textbf {\bibinfo {volume}
  {47}},\ \bibinfo {pages} {835} (\bibinfo {year} {2014})}\BibitemShut
  {NoStop}%
\end{thebibliography}%

\end{document}